\documentclass[5p,,preprint,12pt,twocolumn]{elsarticle}

\usepackage{tabulary,xcolor}
\usepackage{amsfonts,amsmath,amssymb}
\usepackage[T1]{fontenc}
\makeatletter
\let\save@ps@pprintTitle\ps@pprintTitle
\def\ps@pprintTitle{\save@ps@pprintTitle\gdef\@oddfoot{\footnotesize\itshape \null\hfill\today}}
\def\hlinewd#1{%
  \noalign{\ifnum0=`}\fi\hrule \@height #1%
  \futurelet\reserved@a\@xhline}

\AtBeginDocument{\ifNAT@numbers \biboptions{sort&compress}\fi}
\makeatother

\usepackage{ifluatex}
\ifluatex
\usepackage{fontspec}
\defaultfontfeatures{Ligatures=TeX}
\usepackage[]{unicode-math}
\unimathsetup{math-style=TeX}
\else 
\usepackage[utf8]{inputenc}
\fi 
\ifluatex\else\usepackage{stmaryrd}\fi

\usepackage{url,multirow,morefloats,floatflt,cancel,tfrupee}
\makeatletter

\AtBeginDocument{\@ifpackageloaded{textcomp}{}{\usepackage{textcomp}}}
\makeatother
\usepackage{colortbl}
\usepackage{xcolor}
\usepackage{pifont}
\usepackage[nointegrals]{wasysym}
\makeatletter

\def\mcWidth#1{\csname TY@F#1\endcsname+\tabcolsep}

\def\cAlignHack{\rightskip\@flushglue\leftskip\@flushglue\parindent\z@\parfillskip\z@skip}
\def\rAlignHack{\rightskip\z@skip\leftskip\@flushglue \parindent\z@\parfillskip\z@skip}

\if@twocolumn\usepackage{dblfloatfix}\fi 
\AtBeginDocument{
\expandafter\ifx\csname eqalign\endcsname\relax
\def\eqalign#1{\null\vcenter{\def\\{\cr}\openup\jot\m@th
  \ialign{\strut$\displaystyle{##}$\hfil&$\displaystyle{{}##}$\hfil
      \crcr#1\crcr}}\,}
\fi
}

\let\lt=<
\let\gt=>
\def\processVert{\ifmmode|\else\textbar\fi}

\@ifundefined{subparagraph}{
\def\subparagraph{\@startsection{paragraph}{5}{2\parindent}{0ex plus 0.1ex minus 0.1ex}%
{0ex}{\normalfont\small\itshape}}%
}{}

\newcommand\role[1]{\unskip}
\newcommand\aucollab[1]{\unskip}
  
\@ifundefined{tsGraphicsScaleX}{\gdef\tsGraphicsScaleX{1}}{}
\@ifundefined{tsGraphicsScaleY}{\gdef\tsGraphicsScaleY{.9}}{}
\def\checkGraphicsWidth{\ifdim\Gin@nat@width>\linewidth
	\tsGraphicsScaleX\linewidth\else\Gin@nat@width\fi}

\def\checkGraphicsHeight{\ifdim\Gin@nat@height>.9\textheight
	\tsGraphicsScaleY\textheight\else\Gin@nat@height\fi}

\def\fixFloatSize#1{}%\@ifundefined{processdelayedfloats}{\setbox0=\hbox{\includegraphics{#1}}\ifnum\wd0<\columnwidth\relax\renewenvironment{figure*}{\begin{figure}}{\end{figure}}\fi}{}}
\let\ts@includegraphics\includegraphics

\def\inlinegraphic[#1]#2{{\edef\@tempa{#1}\edef\baseline@shift{\ifx\@tempa\@empty0\else#1\fi}\edef\tempZ{\the\numexpr(\numexpr(\baseline@shift*\f@size/100))}\protect\raisebox{\tempZ pt}{\ts@includegraphics{#2}}}}

\AtBeginDocument{\def\includegraphics{\@ifnextchar[{\ts@includegraphics}{\ts@includegraphics[width=\checkGraphicsWidth,height=\checkGraphicsHeight,keepaspectratio]}}}

\def\URL#1#2{\@ifundefined{href}{#2}{\href{#1}{#2}}}

\def\UrlOrds{\do\*\do\-\do\~\do\'\do\"\do\-}%
\g@addto@macro{\UrlBreaks}{\UrlOrds}
\makeatother

\begin{document}

\begin{frontmatter}
	
\title{A guide to emerging technologies for large-scale and whole brain optical imaging of neuronal activity}
    
\author[aff0e4a309581cdbbd45cabd4a3ddb826c0]{Siegfried Weisenburger}
\ead{weisenburger@rockefeller.edu}
\author[aff0e4a309581cdbbd45cabd4a3ddb826c0,aff787bc796c28768532f35b53b107a5e12,affd33be05a3fb48936c72b0cf829cac50f]{Alipasha Vaziri\corref{contrib-8160607a9208bd6d17bf20bbb215d37d}}
\ead{vaziri@rockefeller.edu}\cortext[contrib-8160607a9208bd6d17bf20bbb215d37d]{Corresponding author.}
    
\address[aff0e4a309581cdbbd45cabd4a3ddb826c0]{Laboratory of Neurotechnology and Biophysics\unskip, 
    The Rockefeller University\unskip, New York\unskip, NY\unskip, USA}
  	
\address[aff787bc796c28768532f35b53b107a5e12]{Kavli Neural Systems Institute\unskip, 
    The Rockefeller University\unskip, New York\unskip, NY\unskip, USA}
  	
\address[affd33be05a3fb48936c72b0cf829cac50f]{
    Research Institute of Molecular Pathology\unskip, Vienna\unskip, Austria}

\begin{abstract}
The mammalian brain is a densely interconnected network that consists of millions to billions of neurons. Decoding how information is represented and processed by this neural circuitry requires the ability to capture and manipulate the dynamics of large populations at high speed and resolution over a large area of the brain. While there has been a rapid increase in use of optical approaches in the neuroscience community over the last two decades, most microscopy approaches lack the ability to record the activity of all neurons comprising a functional network across the mammalian brain at relevant temporal and spatial resolution. 

In this review, we survey the recent development in the optical calcium imaging technologies in this regard and provide an overview of the strengths and limitations of each modality and their potential for scalability. We provide a guidance from a biological user perspective that is driven by the typical biological applications and sample conditions. We also discuss the potential for future advances and synergies that could be obtained through hybrid approaches or other modalities.

\end{abstract}
	
\end{frontmatter}
    
\section{\textbf{Introduction}}
One of the key goals of neuroscience is to decipher how the complex dynamics of neuronal activity in the brain is related to perception, cognition and behavior, and what the underlying principles for information processing by these circuits are (Averbeck et al. 2006). Mammalian brains consist of millions to billions of neurons, each of which making thousands of connections to other neurons (Herculano-Houzel et al. 2006, 2007). This results in an extreme density of interconnectivity in the brain, so that randomly selected neurons are, on average, only via a few neurons connected to each other. From this anatomical fact alone it becomes clear that, in such a densely interconnected system, information can be often represented in a distributed fashion and any given neuron can be involved in representation of multiple stimuli or behavioral states in a context-dependent manner. Progress in identifying the ``neuronal code'' has been limited by the lack of appropriate tools and technologies that would allow to record and manipulate the activity of many {\textemdash} and potentially all {\textemdash} neurons within a functional network. This requires methods that enable: \textit{(1)} control of arbitrary spatiotemporal patterns of neural activity and \textit{(2)} read out of the entire system at adequate temporal and spatial resolution.

One of the first approaches for simultaneous measurement of the activity of multiple neurons has been based on electrodes that capture the electrical responses from neurons within the brain region of interest (Kr{\"{u}}ger 2005). Extracellular neuronal recordings with multi-electrode arrays (Buzsaki 2004) are now used together with algorithms for spike sorting to distinguish signals from different neurons based on their waveform (Harris et al. 2016, Rey et al. 2015). However, insertion of electrodes is invasive and approaches for spike sorting work only within the immediate vicinity of the electrodes where the peak amplitudes are high. This, depending on the number of electrodes used, currently limits the effective number of neurons that can be interrogated to \textasciitilde 1,000 {\textendash} 2,000 neurons (Berenyi et al. 2014, Hilgen et al. 2017). 

Functional MRI (Ogawa et al. 1990), in contrast, allows for whole-brain recordings in a variety of nervous systems including humans in a non-invasiveness manner. The contrast in fMRI stems from the hemodynamic response, which is based on changes of the blood oxygenation level in response to neuronal activity. Consequently, fMRI is not a direct measure of neuronal activity and provides a temporal resolution that is only in the sub-Hz regime. Moreover, limited by the gradient of magnetic fields that can be created, its spatial resolution is limited to the millimeter range such that each voxel contains millions of neurons (Huettel et al. 2008).

Another approach to record neuronal activity is based on optical microscopy in combination with functional contrast agents. The first experiments based on this approach used purified proteins from jellyfish (Shimomura et al. 1962), and organic fluorescent dye molecules that are voltage sensitive as reporters of membrane depolarization, as well as dyes that can act as reporters of concentration changes of ions such as Ca\ensuremath{^{2+}} (Grinvald et al. 1987, Grynkiewicz et al. 1985, Salzberg et al. 1977). Transformative for the field of optical imaging of neuronal activity was the conception of GECIs (Miyawaki et al. 1997, Nakai et al. 2001). The advent of these novel probes allowed advanced optical microscopy methods such as 2p excitation fluorescence microscopy (Denk et al. 1990, So et al. 2000) to be used to their full potential, allowing to record neuronal activity at cellular to sub-cellular resolution at tissue depths up to about 1mm, comparable to the thickness of the mouse cortex (Chen et al. 2013, Mao et al. 2008). In the past decade, the number of new tools and technologies has soared, with an exponential increase over the last five years, in part fueled by interdisciplinary ventures such as the Brain Research through Advancing Innovative Neurotechnologies (BRAIN) Initiative (Insel et al. 2013) as well as other initiatives (Yuste \& Bargmann 2017). Today, whole-brain functional imaging of small organisms such as \textit{C. elegans} (Schr{\"{o}}del et al. 2013) or zebrafish larvae (Ahrens et al. 2013, Prevedel et al. 2014) have been demonstrated, and current developments are aimed at large-scale recording and interrogation of mammalian brains with ever-increasing neuronal population size.

The aim of this review is to provide an overview of the various microscopy techniques for Ca\ensuremath{^{2+}} imaging, that are available today, from a conceptual point of view while providing guidelines for biological users which allow them to identify the most-suited techniques for their particular biological questions and context. While we have limited ourselves to only optical methods for Ca\ensuremath{^{2+}} imaging in this review, a number of the discussed tradeoffs also apply to optogenetics (Boyden et al. 2005, Nagel et al. 2002, Vaziri \& Emiliani 2012, Zhang et al. 2007) and optical voltage imaging (Xu et al. 2017).
    
\section{\textbf{General microscopy concepts and considerations}}
To gain access to the full level of information that a brain represents at any given moment in time, a method capable of capturing the activity of every single neuron with adequate temporal and synaptic resolution over the entire brain would be required. It would need to allow for non-invasive recording in mammalian brains over extended periods of time, typically on the order of hours as needed for behavioral studies, and for a number of sessions over several weeks to months. Figure 1 gives an overview of some numerical values of the key anatomical features for the brains of traditionally used model systems, and it illustrates the order of magnitude differences in neuron numbers, synapses or brain volume, that imaging methods must cover. While none of the currently available techniques is able to achieve these requirements at the same time, a plethora of tools, especially various microscopy approaches, have emerged over the recent years aimed at addressing this grand challenge at different levels. In particular, volumetric microscopy methods have been shown to achieve single cell resolution {\textemdash} and in some cases synaptic resolution {\textemdash} at acquisition rates comparable to or faster than the average rate of activity of cortical neurons.

\bgroup
\fixFloatSize{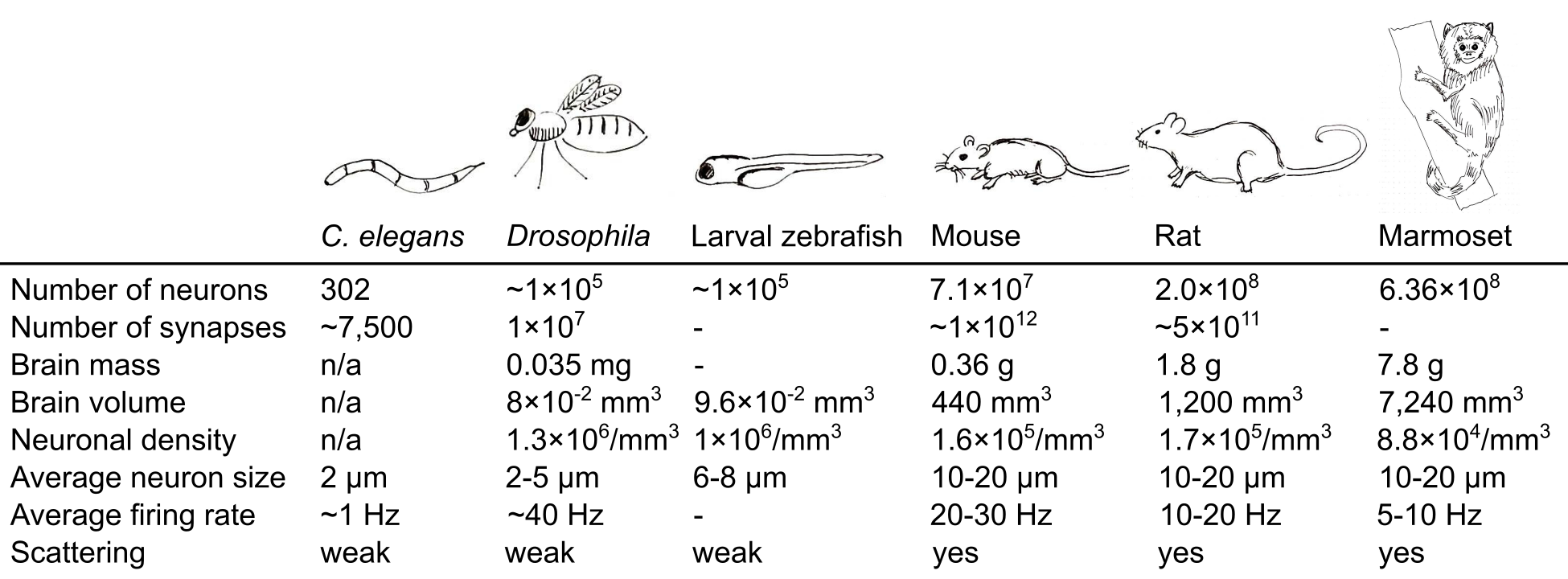}
\begin{figure*}[!htbp]
\centering \makeatletter\IfFileExists{94d76441-0a3b-4ec6-9acb-04fe5793831c-ufig1_organisms.png}{\includegraphics{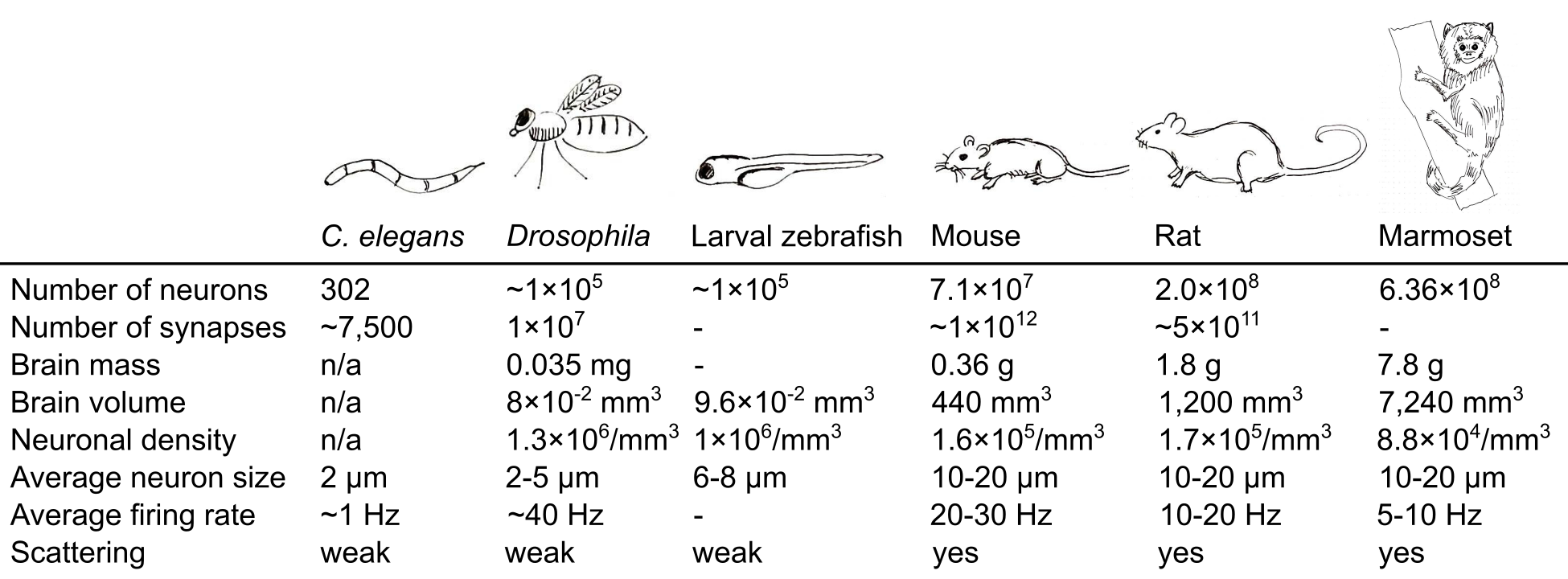}}{}
\makeatother 
\caption{{Overview of numerical values for the key anatomical parameters of the nervous systems /brains of traditionally used model organisms in neuroscience. The table illustrates the order of magnitude differences in neuron numbers, synapses, and brain volumes that imaging methods need to cover. Numbers based on (White et al. 1986): C. elegans,(Naumann et al. 2010): zebrafish larvae, (Gouwens \& Wilson 2009, Rein et al. 2002): drosophila,(Herculano-Houzel et al. 2006, Howarth et al.2012, Kovacevic 2004, Simpson 2009): mouse, (Braitenberg \& Sch{\"{u}}z 1998, Buzsaki \&Mizuseki 2014, Herculano-Houzel et al. 2006, Martin et al. 2010, Sahin et al.2001): rat, (Buzsaki \& Mizuseki 2014, Herculano-Houzelet al. 2007): marmoset.}}
\label{figure-470d9a13d7fcc8d2a2cf0dddfa6cceb8}
\end{figure*}
\egroup
The anatomical properties (Fig. 1) define the required spatial and temporal resolution, however, they also establish the sparsity in time and space, within which methods need to operate, but also what they can actively exploit to relax some of the optical constraints. Key parameters of an imaging modality are its spatial resolution, the acquisition speed and the volume size it can record from. These parameters are linked and further determined by biological and technical limitations: When imaging in a transparent medium, the NA and the wavelength \textit{\ensuremath{\lambda } }determine the spatial resolution, \textit{d } = \textit{\ensuremath{\lambda }}/(2NA) (Weisenburger \& Sandoghdar 2015). In diffraction-limited imaging, the spatial resolution \textit{d}, the size of the acquisition volume \textit{V}, and the number of acquired voxels \textit{n} per time unit determine the volume rate. Neglecting any overhead due to technical limitations, the number of acquired voxels per second is \textit{n} = \textit{M}/\textit{\ensuremath{\Delta }t}, where \textit{\ensuremath{\Delta }t} is the voxel dwell time and \textit{M} is the multiplicity of the acquisition, that is the number of voxels recorded in parallel. To avoid acquisition crosstalk and obtain optimal signal per voxel in fluorescence imaging, the dwell time \textit{\ensuremath{\Delta }t} must be longer than the fluorescence lifetime, which is the time the fluorophore spends in the excited state before emitting a fluorescence photon, typically on the order of a few nanoseconds. Besides this fundamental limit, properties of the sample and the used fluorophore, such as fluorophore concentration and quantum yield, limit the dwell time in a more practical fashion in the required SNR.

A limiting factor for all imaging modalities is photodamage: One source of damage is related to the average amount of power that is deposited into the tissue causing tissue heating. Onset of damage has been observed at temperatures above 40\ensuremath{^\circ}C in the mammalian brain (Podgorski \& Ranganathan 2016). At the same time it has been shown that cranial windows and immersion water can lower the superficial brain temperature to 32\ensuremath{^\circ}C (Kalmbach \& Waters 2012). The maximum average laser power for which heating will be non-detrimental depends on the illuminated volume, the size of the window and imaging depth, as well as the thermal properties of the brain tissue and the vasculature. For 2p in vivo imaging in the mouse brain, average laser powers of about 250 mW have been used without observation of damage as reported by immunohistochemistry (Podgorski \& Ranganathan 2016, Prevedel et al. 2016). Another possible source of damage is nonlinear photo-damage which is caused by nonlinear tissue absorption, intracellular dielectric breakdown or possible plasma formation due to high peak intensities (Hopt \& Neher 2001, Koester et al. 1999). Reported damage thresholds for non-linear photo-damage are \textasciitilde 20 nJ/\ensuremath{\mu }m\ensuremath{^{2}} (Hopt \& Neher 2001), which corresponds to \textasciitilde 10 mW in a diffraction-limited focus of a standard 80-MHz titanium sapphire oscillator. Any limitations in the employable average power or peak intensity affects the available signal or SNR and has, in turn, an influence on reachable depth or acquisition speed. 

Since the discussed tissue heating and photodamage mechanisms restrict the maximum average power and peak intensities that the issue can tolerate, they also indirectly restrict the multiplicity \textit{M, } the number of voxels that are simultaneously recorded at each time point. Furthermore, when imaging in scattering tissue while using a 2D detector array such as a camera, the ability to distinguish nearby voxels degrades with increasing depth due to scatter-induced crosstalk between neighboring voxels, setting another limit on the practically utilizable multiplicity \textit{M}. These sets of relations and interdependencies between the fundamental imaging parameters and conditions, within which samples can be imaged, shows that there is no single ideal method. Each system or biological question brings a different set of conditions, challenges and opportunities.
    
\section{\textbf{Molecular indicators for optical readout of neuronal activity}}
Fluorescence is a key contrast mechanism in biological imaging due to the diversity of labeling strategies that it offers, its high contrast, sensitivity, and the possibility to obtain cell type specific labeling when genetically expressible fluorescence labels are used. In the case of functional imaging, neuronal activity is either directly or indirectly translated into the modulation of a fluorescence signal by a molecular sensor. A dramatic advance for functional fluorescence imaging was the advent of GECIs in the beginning of the 2000s (Miyawaki et al. 1997, Nakai et al. 2001). GECIs revolutionized functional neuronal imaging for several reasons: The most recent generation of GECIs provide superb SBR (Chen et al. 2013) which has enabled recording of neuronal activity at several hundreds of microns in the scattering rodent tissue in vivo. Additionally, they offer genetic targeting, allowing for recording of activity from specific neuronal cell types. Finally, they also allow for a diverse range of delivery strategies. Viral transduction strategies through e.g. lenti- and adeno-associated viruses (Dittgen et al. 2004, Tian et al. 2009) including systemic delivery approaches (Chan et al. 2017, Foust et al. 2008, Shimogori \& Ogawa 2008), as well as methods for genomic integration of the transgenes and generation of transgenic animals (Ji et al. 2004, Tallini et al. 2006, Zariwala et al. 2012) are now being routinely used (Grienberger \& Konnerth 2012).

The Ca\ensuremath{^{2+}} reporter GCaMP was created by fusing a GFP to calmodulin, a Ca\ensuremath{^{2+}} binding protein, such that a conformational change modifies the chemical environment of the chromophore upon Ca\ensuremath{^{2+}} binding. Thus, neurons expressing GCaMP show a weak fluorescence signal when they are inactive and become brightly fluorescent upon onset of neuronal activity due to the influx of Ca\ensuremath{^{2+}} ions. Ca\ensuremath{^{2+}} reporters have been continuously improved in brightness, response time, fluorescence modulation or binding affinity (Akerboom et al. 2012, Chen et al. 2013, Nagai et al. 2004, Tian et al. 2009). The currently widely used version GCaMP6f is one of the fastest Ca\ensuremath{^{2+}} indicators with a decay time of \textasciitilde 140 ms and is still very bright with a strong fluorescence modulation of \textasciitilde 20 \% for a single action potential (Chen et al. 2013). 

All Ca\ensuremath{^{2+}} imaging techniques share some fundamental limitations due to the fact that they are a second messenger for the membrane depolarization (Lin \& Schnitzer 2016). The time scale of the fluorescent signal in Ca\ensuremath{^{2+}} imaging is dictated by diffusion, Ca\ensuremath{^{2+}} buffering and cooperativity. Their reaction is about two orders of magnitude slower than the time scale of an action potential which results in inaccuracies and variabilities in spike timing and response characteristics. The Ca\ensuremath{^{2+}} signal also saturates during bursts of activity, limiting the maximal fluorescence modulation to \textasciitilde 200 \%. Furthermore, GECIs typically cannot report any membrane hyperpolarization or sub-threshold voltage changes (Lin \& Schnitzer 2016). 

Thus, parallel efforts are aimed at the development of genetically encoded voltage indicators (GEVIs) as a direct reporter of the membrane polarization dynamics. Currently, GEVIs do not provide the same activity contrast ratios as GECIs, but they are able to capture neuronal signals with \textasciitilde 10 ms time resolution (Chamberland et al. 2017, Hochbaum et al. 2014). We expect that their further development will lead to a synergetic impact with the technological progress on the microscopy side as discussed in this review.
    
\section{\textbf{Computational signal extraction and computational imaging}}
The steep increase in computer processing power over the past decades enabled the emergence of computational imaging methods where computation becomes an integral part of the imaging process. In a range of computational imaging strategies, any form of prior knowledge about the sample or the experimental instrumentation is used to enable or enhance the capabilities of the imaging system. Examples include statistical tools such as independent component analysis (ICA) (Mukamel et al. 2009) or non-negative matrix factorization (NMF) techniques (Pnevmatikakis et al. 2016) that have been shown to allow the extraction and de-mixing of neuronal signals even when they are overlapping in space or time. While ICA is a linear de-mixing method, nonlinear, NMF-based techniques allow for a better de-mixing in the case of strongly overlapping sources (Pnevmatikakis et al. 2016). NMF makes use of the fact that spatiotemporal activity can be approximated as the product of a spatial matrix containing the location of each neuron and a temporal matrix representing their activity. When seeded or informed by prior knowledge about the location of neurons, such an approach can de-mix neuronal signals stemming from deep inside scattering tissue even when combined with wide-field detection, as recently demonstrated in combination with LFM by the technique of seeded iterative de-mixing (SID) (N{\"{o}}bauer et al. 2017). 

In general, computational deconvolution and reconstruction methods can reduce the burden on the optics or generally the demands on instrumentation (Friedrich et al. 2017). Extrapolating from the current trends that point towards an ever-increasing complexity of the imaging apparatus, the use of computational imaging and other tools to relax the constraints on the instrumentation are expected to become increasingly important and an integral aspect of future optical engineering efforts.
    
\section{\textbf{Categorization of current microscopy methods used for Ca\ensuremath{^{2+}} imaging}}
Current techniques for Ca\ensuremath{^{2+ }}imaging can be broadly categorized according to their degree of parallelization of acquisition and whether the acquisition is unbiased or prior-based as the obtainable acquisition speeds and the applicability of the various methods to different scattering properties of the sample are determined by them (Fig. 2). In addition, there are currently several strategies and emerging techniques for optical recording at greater depths, which we will discuss in this section and describe their advantages and limitations.

\bgroup
\fixFloatSize{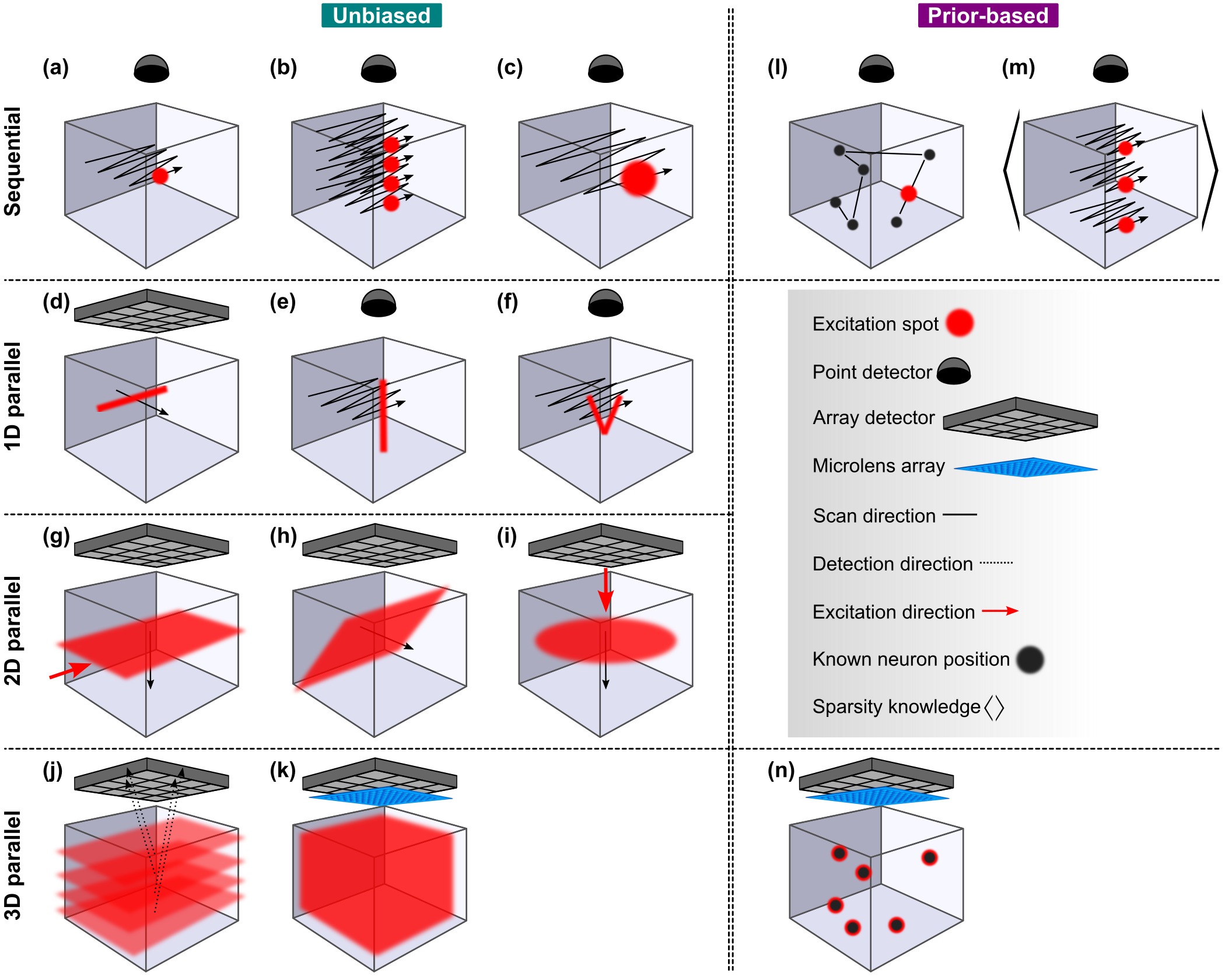}
\begin{figure*}[!htbp]
\centering \makeatletter\IfFileExists{764a4d0a-b8b3-4ac4-9d16-322caaf367cc-ufig2_modalities.png}{\includegraphics{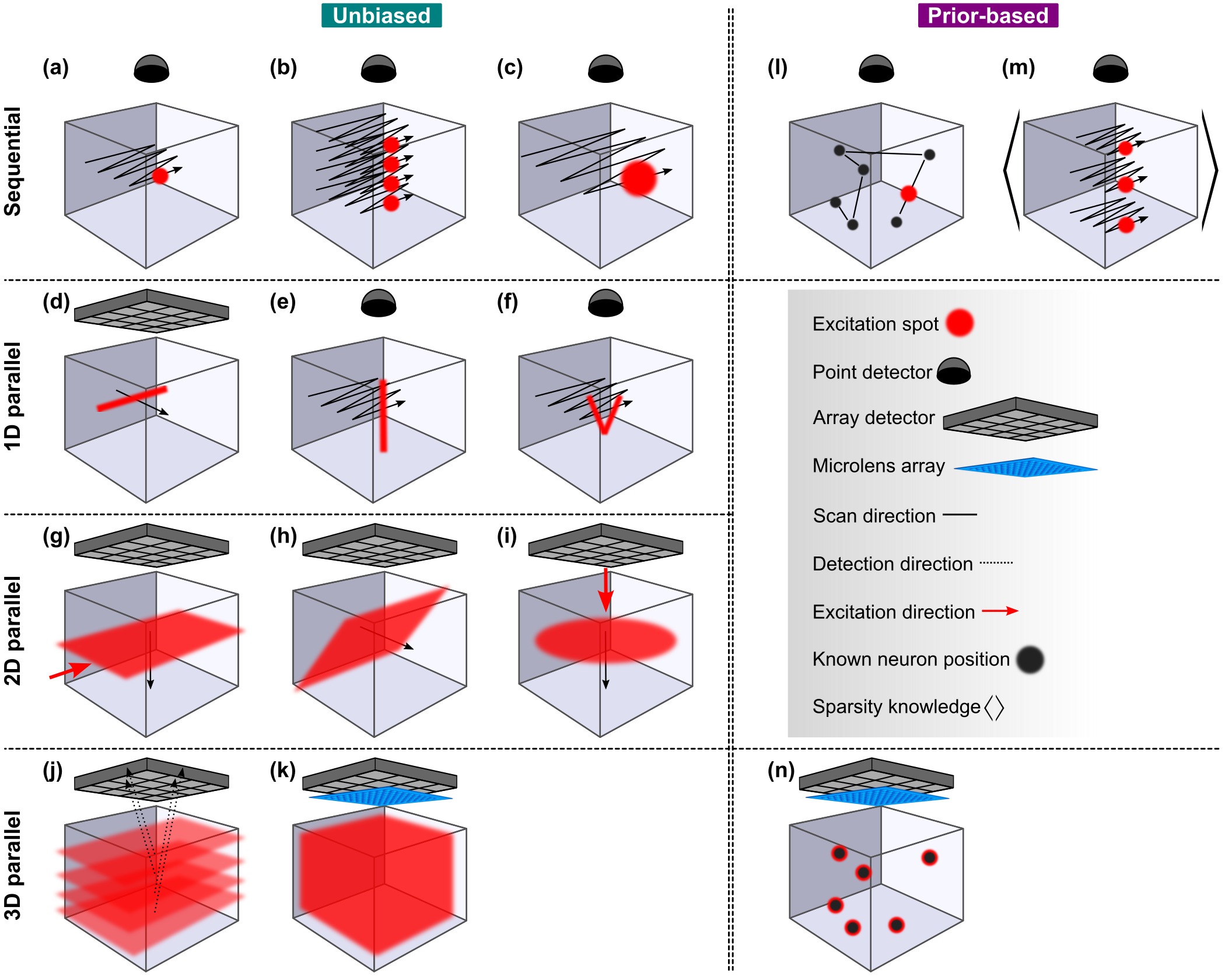}}{}
\makeatother 
\caption{{Illustration of imaging modalities, categorized based on the acquisition mode (fully sequential, 1D parallel, 2D parallel and 3Dparallel) and sampling strategy (unbiased /prior-based sampling). Excitation and detection is in epi-configuration from the top if not indicated otherwise. See legend for an explanation of the used symbols.}}
\label{figure-9566236f5a065f9006d7f648400b25f6}
\end{figure*}
\egroup

\subsection{\textit{\textbf{Mode of acquisition: Parallel versus sequential}}}In a sequential acquisition, a focal spot is scanned across one or more dimensions in space to cover the entire volume. In a parallel acquisition mode, a certain number or all voxels are recorded simultaneously. 

Based on this definition, a range of techniques exists that share some degree of parallelization with some degree of sequential acquisition (Fig 2). In fully sequential techniques such as diffraction-limited point scanning, the location of the signal is determined by the instantaneous position of the excitation beam. Consequently, a point detector can be used that collects emitted fluorescence photons irrespective of the path on which they reach the detector. Thus, fully sequential acquisition techniques, and to a lesser extent partially sequential techniques, provide robustness towards scattering and are therefore well-suited for deep tissue imaging. They typically also deposit only a small amount of energy per time unit in the sample which reduces or avoids the risk of tissue damage. Yet, this comes at the cost of a reduced time resolution. A parallelized acquisition mode based on a detector array or a camera allows for higher volume rates, but, depending on the scattering properties of the tissue, at the cost of potential scatter-induced crosstalk between neighboring camera pixels. Furthermore, the deposited average power and the potential risk for tissue damage increases with the degree of parallelization.

\bgroup
\fixFloatSize{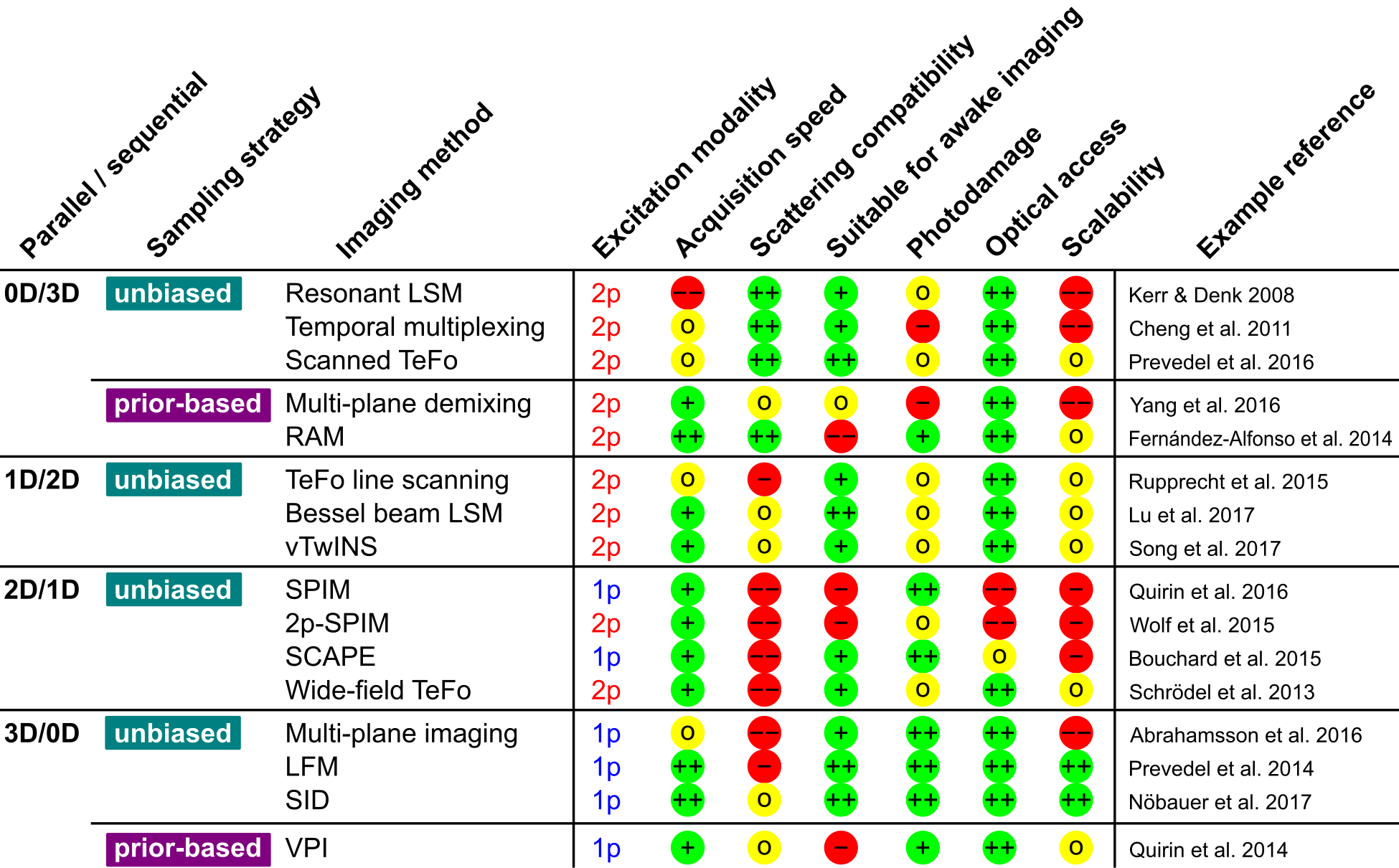}
\begin{figure*}[!htbp]
\centering \makeatletter\IfFileExists{cdcbcd4c-fbca-41ba-a826-3cb18c520b39-ufig3_methods.png}{\includegraphics{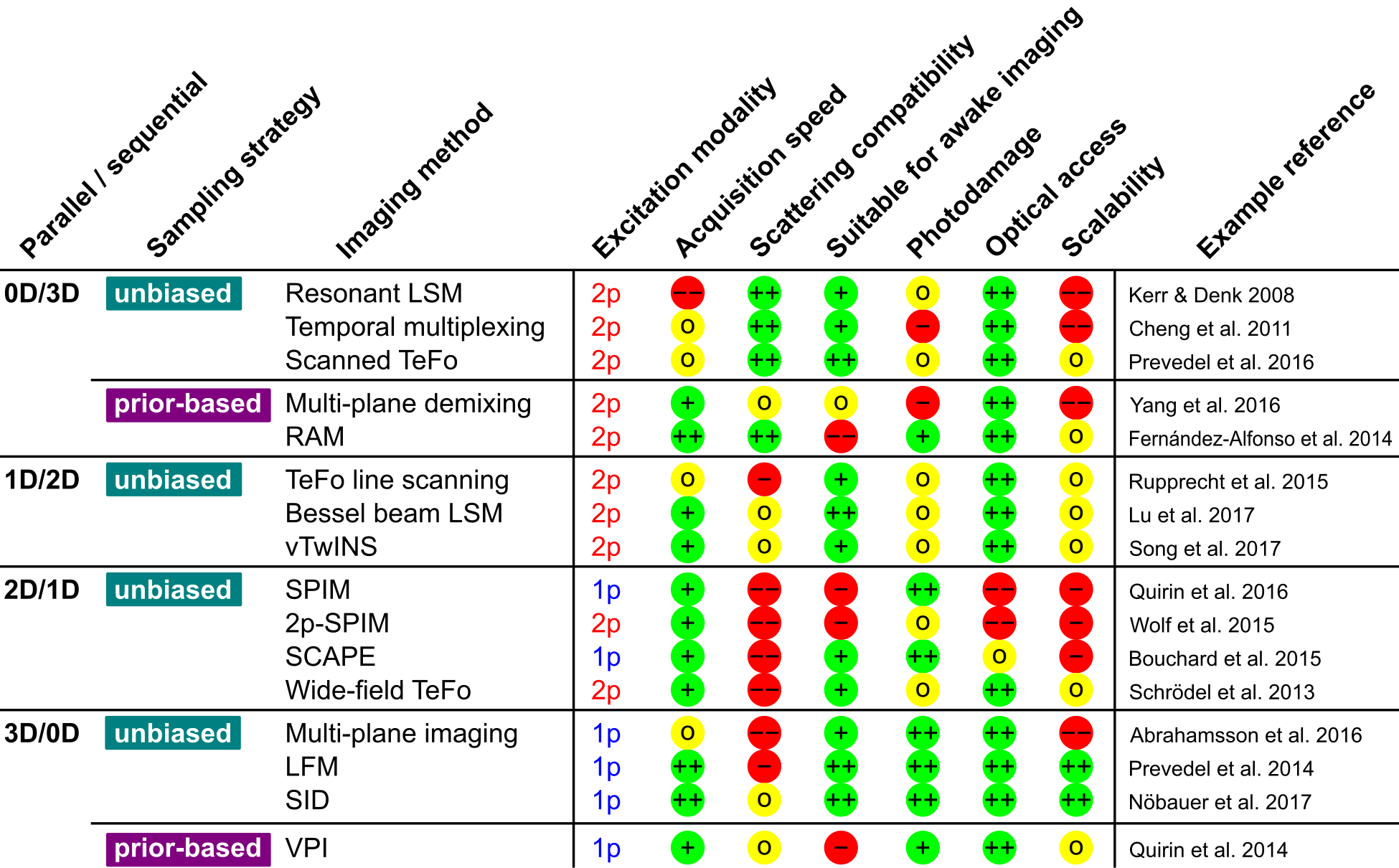}}{}
\makeatother 
\caption{{Overview of the categories and key imaging properties of the available techniques for functional optical imaging, and their strengths and limitations. The modalities are categorized based on the acquisition mode(fully sequential, 1D parallel, 2D parallel and 3D parallel) and sampling strategy (unbiased/prior-based sampling). The key imaging properties, acquisition speed, scattering compatibility, suitability for awake imaging(including susceptibility to motion artifacts), photodamage, optical access, and scalability of the method, are rated with ++ (best, green), + (green), 0 (yellow), \ensuremath{-} (red), \ensuremath{-}\ensuremath{-} (worst, red). Example references for the techniques are provided.}}
\label{figure-e9fc726f66aaaa3cafa923999e399837}
\end{figure*}
\egroup

\subsubsection{\textit{Sequential acquisition:}}The most basic fully sequential technique for 3D imaging using point scanning is confocal LSM (Pawley 2006), Fig. 2a. Confocal microscopy is rarely used for \textit{in vivo} imaging in scattering brains due its low efficiency to collect emitted fluorescence photons and level of fluorophore bleaching. As a result, the application of some of its variants such as confocal spinning disk microscopy has been limited to small and semi-transparent organisms such as \textit{C. elegans} (Kato et al. 2015).

A prominent technique for functional imaging using point scanning is 2p LSM (Denk et al. 1990, So et al. 2000). Here, the absorption of two lower-energy photons is necessary for the fluorophore to be able to emit fluorescence. This means that the 2p absorption probability depends on the light intensity in the focal spot squared. To reach the necessary peak intensities for 2p fluorescence excitation, pulsed femtosecond lasers are typically utilized. The nonlinearity in the excitation comes with another benefit: Only fluorescent molecules in the laser focus experience a sufficient intensity for 2p excitation and subsequent fluorescence emission. This results in a significant reduction of the out-of-focus fluorescence and thereby an increase in the SNR and SBR while minimizing photobleaching outside of the focus. 2p excitation is also beneficial for a deeper penetration of the excitation light owing to the lower absorption of tissue in the NIR region of the spectrum and because the scattering intensity approximated by Mie scattering scales as \textasciitilde 1/\textit{\ensuremath{\lambda }}\ensuremath{^{4}}. An additional design consideration when using ultrafast laser pulses for excitation in both scattering and transparent tissue is pulse broadening due to dispersion which reduces the fluorophore excitation probability and therefore the SNR (Trebino \& Zeek 2000).

Although 2p microscopy has enabled a wide range of studies in scattering brain tissue, primarily in rodents, during the past two decades, its low speed has limited neuroscience studies to planar recordings, even with fast resonant scanning schemes (Kerr \& Denk 2008). 

One way to increase the speed is to introduce multiplicity in the acquisition. This can be done by scanning across the volume with multiple beams simultaneously, a technique termed temporal multiplexing (Amir et al. 2007, Cheng et al. 2011, Stirman et al. 2016). Here, the excitation laser beam is divided into multiple beamlets, which are delayed with respect to each other, and then are used for simultaneous excitation and recording in different lateral or axial positions in the imaging volume (Fig. 2b). While a single point detector such as a PMT is used, it is possible in this scheme to distinguish the fluorescence signal from different spatial regions by their different arrival times at the PMT. Four-times multiplexing schemes have been demonstrated, in which the degree of multiplicity was limited by the typical 80 MHz laser repetition rate of a titanium sapphire oscillator, corresponding to a 12.5 ns inter-pulse interval, and the fluorescence lifetime which is on the order of a few nanoseconds (Cheng et al. 2011). Using this approach, multi-plane diffraction-limited imaging over a FOV of 400 x 400 \ensuremath{\mu }m has been shown for four planes at a rate of 60 fps over an axial range of less than 100 \ensuremath{\mu }m (Cheng et al. 2011). In the above scheme, the required average power scales linearly with the number of multiplexed beams, so that photodamage due to tissue heating ultimately limits the multiplicity that can be practically achieved using this technique. This method is also not easily scalable as the complexity of the optical setup increases with the multiplicity.

\bgroup
\fixFloatSize{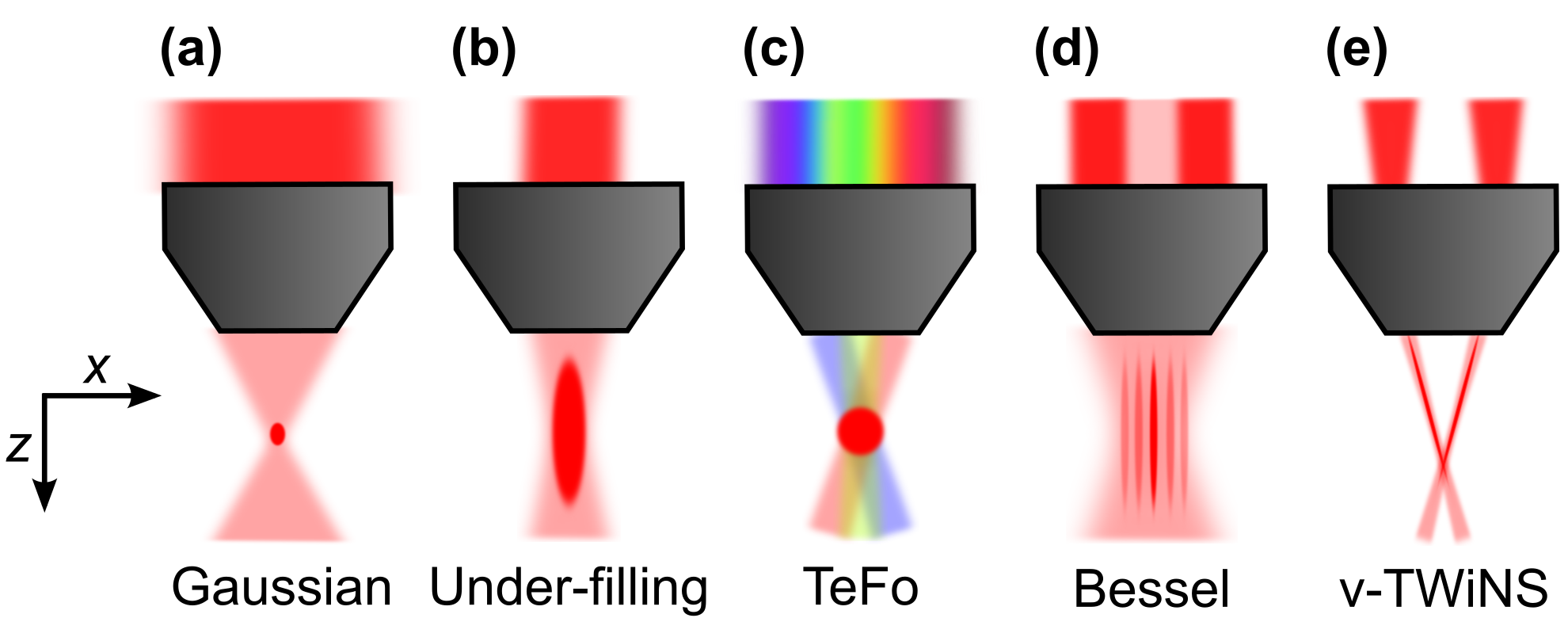}
\begin{figure}[!htbp]
\centering \makeatletter\IfFileExists{b9404996-9939-4f3c-808c-35539716572a-ufig4_psf.png}{\includegraphics{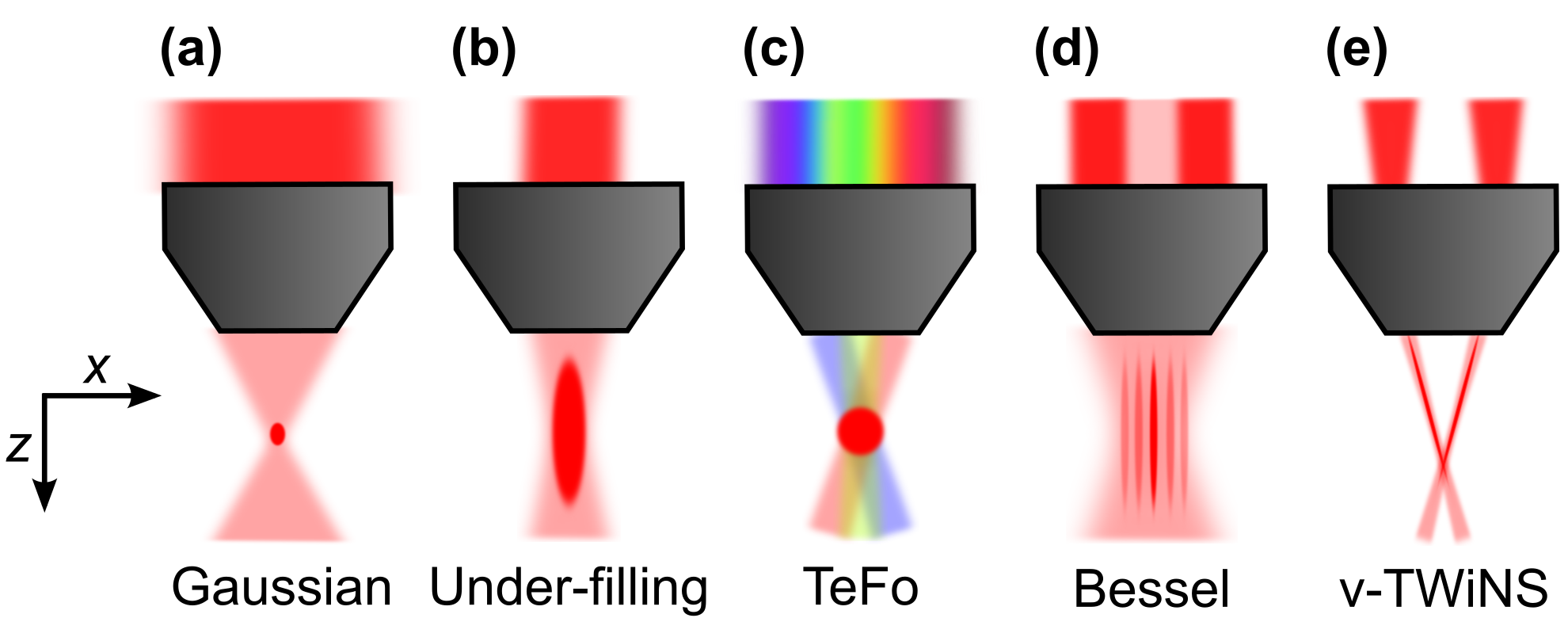}}{}
\makeatother 
\caption{{Overview of excitation modalities and their corresponding PSF of different imaging method: (a) Gaussian focus, (b) under-filled Gaussian focus, (c) Isotropic focus with light sculpting using temporal focusing, (d) Bessel beam generated by an annular pattern in the back aperture, (e) V-shaped PSF.}}
\label{figure-511f752ea099606d7b8d6de3366d6f08}
\end{figure}
\egroup
Another approach to increase the volumetric recording speed in sequential imaging is to optimize and adapt the spatial resolution and sampling frequency to the structure of interest (Fig. 2c). Since the average size of neuronal cell bodies in mammalian cortex (\textasciitilde 10-20 \ensuremath{\mu }m) is more than ten times larger in diameter than a diffraction-limited focus, spatial resolution can be traded for speed, allowing recording of neuronal signals from larger populations at a higher speed while maintaining single-cell resolution, and depending on excitation depth can allow for a higher SNR at the same average power. Fundamental constraints from optics do not allow an arbitrarily shaped excitation PSF when a Gaussian beam is used. In particular, isotropic resolution cannot be achieved with a Gaussian beam because of the intrinsic coupling of the lateral (\textit{d}) and the axial (\textit{z}) confinement of excitation in a Gaussian focus, \textit{z} \textasciitilde  \textit{d}\ensuremath{^{2}} (Fig. 4a,b). Using light sculpting based on TeFo (Oron et al. 2005, Zhu et al. 2005), this issue can be addressed: The spectrum of a femtosecond laser pulse is spatially dispersed by using a grating. Imaging the spot on the grating onto the sample with a telescope that consist of the microscope objective and a lens, results in a configuration where the spectral components of the laser pulse overlap in time and space only within a small axial region in which 2p excitation can occur. Thereby, axial sectioning can be achieved independently from the lateral size of the excitation spot through dispersion and, thus, an effective de-coupling of the axial and lateral confinement of excitation can be realized (Fig. 4c). In our laboratory, we have used scanned TeFo (s-TeFo) to shape the 3D PSF of our microscope near-isotropically to match the required size to sample neuronal cell bodies in mouse cortex (\textasciitilde 10-15 \ensuremath{\mu }m) (Prevedel et al. 2016). To generate sufficient pulse energy for exciting the \textasciitilde 100 times larger spot, a fiber amplified laser system was used. We have shown that  s-TeFo enabled unbiased single- and dual-plane high-speed (up to 160 Hz) Ca\ensuremath{^{2+}} imaging as well as in vivo volumetric Ca\ensuremath{^{2+}} imaging of a mouse cortical column (0.5 mm x 0.5 mm x 0.5 mm) at single-cell resolution and fast volume rates (3-6 Hz). An advantage of this approach is that the peak intensities are lower because of the larger spot size, which is beneficial for preventing nonlinear tissue damage. On the other hand, the larger spot size requires a higher average power, so that s-TeFo operates closer to limits imposed by heating, even though no detrimental effects due to sample heating as reported through immunohistochemistry were observed under the above conditions. Finally, the reduced spatial resolution can be compensated by using computational signal extraction approaches to reduce possible mixing of neuronal signals (Friedrich et al. 2017, Pnevmatikakis et al. 2016).

\subsubsection{\textit{1D parallel acquisition:}}A possibility to reduce the sequential scanning to two dimensions is scanning a line along the lateral or axial dimension instead of a focal spot across the volume. One way is to generate a line, which is scanned across the volume while the fluorescence signal is being recorded using an array detector, e.g. a camera (Fig. 2d). In this case, crosstalk due to scattering of the fluorescence light from neighboring pixels in the parallel readout limits the spatial resolution and the reachable imaging depth. To improve this, the synchronization of a line-shaped illumination with the rolling shutter readout of modern CMOS cameras can be exploited (Baumgart \& Kubitscheck 2012, Spiecker 2011). Our laboratory has demonstrated the full potential of this method by combining the line-scanning acquisition scheme with light sculpting using TeFo (Rupprecht et al. 2015). We could demonstrate high-speed planar imaging of 200 x 200 \ensuremath{\mu }m FOV at 75 fps down to a depth of 75 \ensuremath{\mu }m in scattering brain tissue. By optimizing the rolling shutter width, scattering effects could be reduced by about a factor of three. 

TeFo based excitation is part of the general category of PSF engineering. Another instantiation of PSF engineering are axially elongated PSFs. They have been shown to enable simultaneous recording along the axial direction and thereby reducing the scanning to a 2D planar fashion (Fig. 2e). An example of such a technique uses Bessel beams which are generated by an annular pattern in the back-focal plane; the resulting interference of plane waves at a shallow angle produces an axially elongated focus (Fig. 4d). Bessel beams have been applied to 2p Ca\ensuremath{^{2+}} imaging in brain slices (Botcherby et al. 2006, Th\'{e}riault et al. 2014) and recently also to volumetric in vivo functional imaging (Lu et al. 2017). In the latter implementation, the neuronal sparsity in the brain volume was exploited to ensure minimal structural overlap in the 2D projection. Neural activity could be monitored at volume rates equivalent to the 2D frame rates, achieving up to 30 vol/s for a \textasciitilde 250 x 250 x 100 \ensuremath{\mu }m volume at depths down to \textasciitilde  160 \ensuremath{\mu }m in zebrafish larvae and fruit fly, as well as scattering mouse and ferret cortex. Bessel foci have a stronger lateral confinement than Gaussian beams in the central peak for the same NA {\textemdash} but the redistribution of power into the side rings requires higher average power, in practice about 3-4 times more in the case of 2p excitation (Lu et al. 2017), and the fraction of power in the side rings becomes larger with increasing NA. This is one of the limitations of this approach as it leads to a high fluorescence background when a one-photon excitation strategy is used and to an increased level of tissue heating in 2p excitation. Furthermore, both practical limitations in the achievable aspect ratio of the focus and the increased mixing of neuronal signals from different planes, as the Bessel beam's axial length is increasing, restrict the accessible axial range of this approach.

An interesting approach where a different type of engineered PSF was used is vTwINS (Song et al. 2017). Here, a volume was imaged using an elongated, V-shaped PSF (Fig. 2f). Such a PSF can be generated by impinging two small converging beams displaced from the optical axis onto the back-focal plane of the microscope objective (Fig. 4e). Due to the V-shaped focus, single neurons appear as spatially separated pairs in the resulting projection where the distance between the projections encodes for the axial position in the volume. Thereby, volumetric recordings from 550 x 550 x 50 \ensuremath{\mu }m at a speed of 30 vol/s could be demonstrated. A disadvantage of this technique is the increased background due to the V-shaped PSF. Moreover, in this technique, the extent of the axial range is limited as the success of decoding axial positions from the projection images relies on sparsity, and thus restricts its application to superficial volumes.

\subsubsection{\textit{2D parallel acquisition:}}To further reduce scanning, 2D parallel acquisition schemes record 2D images using a camera while illuminating only a thin slice of the sample volume orthogonally to the direction of observation (Fig. 2g) (Huisken et al. 2004, Voie et al. 1993). These techniques are generally referred to as SPIM or `light sheet' microscopy while there are also other names for the different realizations of this general idea (Santi 2011). SPIM comes in various flavors that differ in the exact optical arrangement, how the light sheet is generated and how the signal is acquired. It has been used to record the activity of the entire brain volume of zebrafish larvae at \textasciitilde 1 Hz, capturing the activity of the majority of all neurons at single-cell resolution (Ahrens et al. 2013). The latest versions of SPIM demonstrated Ca\ensuremath{^{2+}} imaging of the zebrafish brain with 420 x 830 x 160 \ensuremath{\mu }m FOV at 33 vol/s and cellular resolution which is more than an order-of-magnitude improvement in volume rate (Quirin et al. 2016). The light sheet is typically generated using a cylindrical lens. While the lateral resolution is determined by the wide-field detection optics, the axial resolution is dictated by the thickness of the light sheet. However, a thin, high-NA light sheet diverges rapidly limiting the FOV. An alternative is to generate the light sheet using a scanning Bessel beam; 2 \ensuremath{\mu }m axial resolution over a span of \textasciitilde 600 \ensuremath{\mu }m has been demonstrated in the context of vasculature imaging in zebrafish (Zhao et al. 2014). There are implementations of SPIM with both one-photon and 2p excitation; an advantage of SPIM with one-photon excitation is that the required excitation powers are much lower compared to 2p excitation. However, one-photon excitation is limited to non- or weakly scattering tissue because the scattering of the illumination beam strongly degrades the illumination light sheet and impairs the spatial resolution. The combination of 2p excitation and SPIM allowed an increased SNR for imaging in weakly scattering tissue (Truong et al. 2011, Wolf et al. 2015). Another restriction is the fact that SPIM requires optical access from at least two orthogonal sides, in the case of opposing light sheets even more (Lemon et al. 2015), to project the excitation light sheet into the sample. The dual-objective geometry limits biological applications to organisms such as zebrafish larvae where optical access from different sites is possible, and the organism needs to be immobilized making it incompatible with behavioral studies.

Another technique that is based on the concept of SPIM but requires optical access from only one direction is swept confocally-aligned planar excitation (SCAPE) (Bouchard et al. 2015, Kumar et al. 2011). Here, an angled, swept light sheet is used for excitation, and the detection is through the same single microscope objective (Fig. 2h). A confocal de-scanning and rotation mapping is used to capture the scanned plane with a camera achieving a volume rate of 10 vol/s for a FOV of 600 x 650 x 135 \ensuremath{\mu }m in scattering mouse brain. As a one-photon modality, also SCAPE has a limited penetration depth in scattering tissue.

One way to overcome the constrained sample geometry of conventional LSM is to use 2p excitation and light sculpting to introduce optical sectioning for wide-field fluorescence imaging in epi-configuration (Fig. 2i). Our laboratory demonstrated wide-field TeFo where a disc-shaped excitation area of \textasciitilde 75 \ensuremath{\mu }m diameter with an axial confinement of \textasciitilde 2 \ensuremath{\mu }m was generated, and then axially scanned (Schr{\"{o}}del et al. 2013). Using this approach, we could for the first time demonstrate whole-brain volumetric imaging in \textit{C. elegans} by recording from 75 x 75 x 40 \ensuremath{\mu }m at 13 vol/s. A 10 kHz-repetition rate laser system was used to provide the required higher peak pulse energies of \textasciitilde 2 \ensuremath{\mu }J at the sample for the wide-field excitation (20 mW average power). Since the required average laser power scales linearly with the excitation area, ultimately damage due to heating poses a limitation to this technique.

\subsubsection{\textit{3D parallel acquisition:}}A fully parallel acquisition modality captures an entire volume in a single acquisition. This requires the ability to map simultaneously all axial information onto a single lateral plane that can then be imaged in a parallel fashion.

One way to realize that experimentally is to image different focal planes simultaneously using different cameras or different regions of a single camera, which has been applied to functional imaging in \textit{C. elegans} using nine focal planes (Abrahamsson et al. 2016) (Fig. 2j). In this one-photon excitation realization, a volume of 40 x 40 x 18 \ensuremath{\mu }m was imaged at 3 vol/s and diffraction-limited resolution. Custom diffractive optical elements are required to split the focal planes and to correct for spherical and chromatic aberrations. While this approach does not require any computational reconstruction, the increasing complexity of the optical system with more focal planes puts a practical limitation on the accessible volumes. A major limitation of this type of multifocal microscopy is the lack of optical sectioning; as a consequence, both SNR and SBR are compromised by out-of-focus fluorescence and because the fluorescence signal is split over the number of focal planes.

Another technique for simultaneous 3D imaging is LFM, which captures both the 2D location and the 2D angular information of incident light at the same time (Levoy et al. 2006, Lippmann 1908). A great advantage of LFM is that it is fully scalable since the volume rate is fundamentally independent of the volume size and only determined by the number of pixels and readout rate of the camera. To realize LFM, a microlens array is placed in the native image plane such that the camera pixels can capture the rays of the light field simultaneously (Fig. 2k). Thus, the PSF encodes for the angular and, in consequence, the axial information, and 3D volumes can be computationally reconstructed using algorithms that solve the inverse problem (Agard 1984, Broxton et al. 2013). In our laboratory, we have shown the application of LFM for whole-brain Ca\ensuremath{^{2+}} imaging in small organisms like \textit{C. elegans} and zebrafish larvae (Prevedel et al. 2014). Recently, other variants of LFM have also been used to demonstrate whole-brain Ca\ensuremath{^{2+}} imaging of freely swimming larval zebrafish (Cong et al. 2017) and \textit{Drosophila} (Aimon et al. 2017). 

4D phase-space measurements, as done by LFM, contain a high redundancy in the information on the 3D location of an object, resulting in a strong robustness against scattering compared to other camera-based wide-field techniques (Liu et al. 2015). By using a computational approach termed Seeded Iterative De-mixing (SID), our laboratory could extend the capabilities of LFM to mammalian cortex and show large-volume functional recording in awake, behaving mice (N{\"{o}}bauer et al. 2017) reaching 30 vol/s at single cell resolution for volumes as large as 900 x 900 x 260 \ensuremath{\mu }m and at depths down to \textasciitilde 380 \ensuremath{\mu }m. It is noteworthy that the signal extraction using SID considerably reduces crosstalk between voxels compared to frame-by-frame image reconstructions. Furthermore, SID reduces the computational cost by about three orders of magnitude. 

An advantage of the one-photon excitation as used in LFM is the low level of required average power minimizing potential tissue damage. A limitation of LFM is its reduced spatial resolution as the microlens array trades lateral resolution for the axial information, albeit potential signal mixing can be addressed by computational approaches. The ultimate depth limit in LFM-based methods is given by multiple scattering events that will lead to a full loss of any directional information of the emitted fluorescence photons and is expected to be in the range of \textasciitilde 400-500 \ensuremath{\mu }m in the cortex.

\subsection{\textit{\textbf{Sampling strategies: Unbiased versus prior-based}}}All imaging modalities discussed above are based on an unbiased sampling strategy. This means that they treat every voxel of the volume in the same way {\textendash} regardless if it contains a neuron or not. If prior knowledge on the specific neuron locations is available, a deterministic, sparse sampling strategy can be used, which will reduce the required samples to the theoretical minimum of one sample per neuron within the volume per time unit. In cases where prior knowledge on only average properties of the sample such as the sparsity is available, the number of spatial samples can be reduced by a probabilistic approach. 

One class of Ca\ensuremath{^{2+}} imaging that exploits spatial sparseness and uses prior knowledge on neuronal positions is RAM (Fern\'{a}ndez-Alfonso et al. 2014, Katona et al. 2012, Reddy et al. 2008). In this modality, fast scanners, such as AODs, are used to access points in 3D that are known to contain neurons or neuronal processes along arbitrary trajectories in the volume (Fig. 2l). By only recording at the a-priori known locations of the neurons, acquisition rates up to \textasciitilde 50 points per kHz can be achieved. A technical issue is that AOD scanners have diffraction efficiencies of \textasciitilde 80-90\%. Since multiple (up to 4) AODs are used for a full 3D RAM, this results in low power efficiency. Moreover, AODs can only be used for small scan angles, typically on the order of a few degrees, limiting the FOV; and they also introduce strong dispersion up to 72,000 fs\ensuremath{^{2}} (Katona et al. 2012). Furthermore, the use of AODs puts limitations on using multiple excitation wavelengths. A fundamental disadvantage of RAM is its higher susceptibility to sample motion which makes its application in awake, behaving animals difficult. 

Another class of techniques using prior spatial information are holographic techniques. They typically generate multiple spots at arbitrary locations within the volume. In one realization, a spatial light modulator in combination with traditional galvanometric scanners has been used to simultaneously image multiple areas or alternatively up to three axially separated layers within the sample (Yang et al. 2016), Fig. 2m. This was done by the spatial light modulator generating multiple beamlets that are then scanned across the FOV via the scanners. The fluorescence signal is detected by a PMT which results in an image that is effectively a super-position of individual images stemming from the different beamlets. To de-multiplex the signals, prior knowledge of the neuron locations, the time-sparsity in their activity as well as a constraint non-negative matrix factorization algorithm can be utilized. In practice, the number of planes in this approach is limited by the available laser power and, more fundamentally, by the neuronal sparsity. Furthermore, the separation between planes needs to be more than \textasciitilde 5 times the size of the axial PSF to avoid decreased optical sectioning and signal mixing. 

Another holographic approach, called volume projection imaging (VPI), uses an SLM to simultaneously excite neurons at locations which are a-priori identified from a conventional 2p LSM image stack (Quirin et al. 2014). Using a custom imaging system that employs a phase mask in the imaging path, the PSF is modified such that the image of each neuron becomes independent of its axial position. Thus, the neurons can be projected onto a single plane without out-of-focus blur of the PSF and recorded on a camera (Fig. 2n). This approach has been used to demonstrate simultaneous 3D imaging within 280 x 270 x 110 \ensuremath{\mu }m at 30 vol/s in zebrafish. As with all techniques relying on spatial priors, difficulties arise with motion during the acquisition in awake, behaving animals. Furthermore, prior knowledge must be available, most likely by using a correlative technique which might not always be possible.

\subsection{\textit{\textbf{Optical recording at greater depth}}}In the case that there is no or only weak scattering, our above considerations regarding spatial resolution and acquisition speed are universally applicable to all sample locations. In the presence of scattering and absorption this is no longer true.

In general, the maximally obtainable imaging depth \textit{z} in optical microscopy is limited by the scattering length \textit{l} in the medium, because the excitation intensity is attenuated as \textit{I} = \textit{I}\ensuremath{_{0}} exp({\textendash}\textit{z}/\textit{l}). In brain tissue, scattering is forward directed with an anisotropy factor g \textasciitilde  0.9, resulting in photons that maintain some directional information even after a few scattering events (Ntziachristos 2010). The scattering length in brain tissue is \textasciitilde 50-100 \ensuremath{\mu }m in the visible part of the spectrum and \textasciitilde 100-400 \ensuremath{\mu }m in the NIR. Since available fluorescent indicators have absorption and emission wavelengths in the visible part of the spectrum, single-photon excitation techniques are limited to a few hundred microns in practice; for certain NIR wavelengths a few millimeters can be reached. In 2p microscopy, the visible indicators can be excited with NIR light which can penetrate deeper into scattering tissue. Secondly, the SBR in 2p microscopy is significantly higher at greater depths since 2p microscopy achieves a high localization of excitation and an effective suppression of out-of-focus fluorescence. 

However, as fewer photons reach the focal volume with increasing tissue depth due to scattering, higher photon densities will be required at the tissue surface to maintain the same level of SBR, which will contribute to the background signal. This will eventually limit the achievable depth in 2p microcopy. One way to achieve high SBR even at greater tissue depths is to use 3p excitation. Here, the higher degree of nonlinearity in the interaction allows for a higher SBR compared to 2p microscopy since the SBR in 3p excitation is given by SBR\ensuremath{_{3p}} \ensuremath{\propto } \textit{z}\ensuremath{^{4}}NA\ensuremath{^{6}}/(\ensuremath{\lambda}\ensuremath{^{3}}\textit{l}) exp({\textendash}\textit{z}/\textit{l}) compared to SBR\ensuremath{_{2p}} \ensuremath{\propto } \textit{z}\ensuremath{^{2}}NA\ensuremath{^{2}}/(\ensuremath{\lambda} \textit{l}) exp({\textendash}\textit{z}/\textit{l}) for 2p excitation (Horton et al. 2013). 3p excitation is also beneficial because the excitation wavelengths for the available green and red Ca\ensuremath{^{2+}} indicators lie within transparency windows of brain tissue where absorption and scattering is reduced, allowing to reach imaging depths beyond one millimeter. Recently, functional imaging of hippocampal neurons at a depth of \textasciitilde 1 mm through the intact mouse cortex has been demonstrated using 3p excitation (Ouzounov et al. 2017). Since the 3p excitation cross-section of fluorophores (\textasciitilde 10\ensuremath{^{\ensuremath{-}}}\ensuremath{^{82}} cm\ensuremath{^{6 }} s\ensuremath{^{2 }} / photon\ensuremath{^{2}}) is considerably smaller than the 2p equivalent (\textasciitilde 10\ensuremath{^{\ensuremath{-}49}} cm\ensuremath{^{4 }} s / photon), the 3p signal is more strongly affected by dispersion and the associated pulse broadening as well as scatter-induced reduction of photon density at the excitation spot (Horton \& Xu 2015). As a result, about one order of magnitude higher pulse energies are necessary in practice to produce a comparable signal level in 3p compared to 2p excitation which, in turn, has to be considered in terms of tissue heating (Wang et al. 2017). 

An alternative approach is to alleviate the underlying issue instead of avoiding it: When enough information about the scattering process is available, one can account for scattering using a corrected wave-front (Vellekoop \& Mosk 2007). Wave front-shaping techniques accomplish this by inverting the corresponding scattering matrix (Conkey et al. 2012). This can be done by using a fluorescent object of a known size, referred to as a guide star, and various types of algorithms that infer the tissue-induced aberrations from the recordings of the guide star and use that information to shape the wave front of the excitation beam such that it counters the effects of aberration at the focal plane. Different methods using various types of algorithms and guide stars have been demonstrated to implement the general idea above (Horstmeyer et al. 2015). This includes photo-acoustic techniques (Lai et al. 2015), the intrinsic 2p fluorescence signal (Katz et al. 2014), or by focus scanning holographic aberration probing (F-SHARP) (Papadopoulos et al. 2016). The required correction patterns for the wave front are usually computed in an iterative fashion and applied to the excitation beam via amplitude or phase wave-front shaping using a spatial light modulator (SLM) or a digital micro-mirror device (DMD). The technical challenge in these methods is that the wave-form correction is only valid for a small FOV within the memory-effect range of the scattering medium, which is several tens of microns in brain tissue. Thus, large FOVs would have to be tiled from individually corrected smaller FOVs. This means that a large number of corrections have to be computed for many sub-volumes, and they also need to be applied fast enough while imaging. Furthermore, in case of in vivo imaging, the correction is only valid for a time period until the scattering channels have de-correlated due to tissue physiology and vasculature or tissue deformation, for which values on the order of 20 minutes have been reported (Papadopoulos et al. 2016). When aberration correction methods are used in combination with Ca\ensuremath{^{2+}} imaging, these effects either impact the volume rate or the volumetric FOV at which neuronal population recording can be practically realized. 
    
\section{\textbf{How to choose a microscopy method?}}
In the following section, we provide guidance for navigating the main microscopy techniques which have emerged over the recent years, and for choosing the suitable method for a particular application. Assuming that the goal is to achieve the highest possible neuronal sampling rate for a given volume size, a maximally parallel acquisition strategy - within the limits set by photo-induced, compatible with the tissue scattering properties - that is only as sequential as necessary, should be used. Neurons or neuronal processes should be spatially sampled as sparsely as possible and as densely as necessary. If we assume that it is desired to access the largest possible volume at the highest possible acquisition speed, imaging modalities with a high degree of parallelization should be chosen if the tissue exhibits no or a low level of scattering. In the presence of scattering and depending on the desired imaging depth, the methods of choice will be dominated by techniques with an increasing level of sequential acquisition.

\bgroup
\fixFloatSize{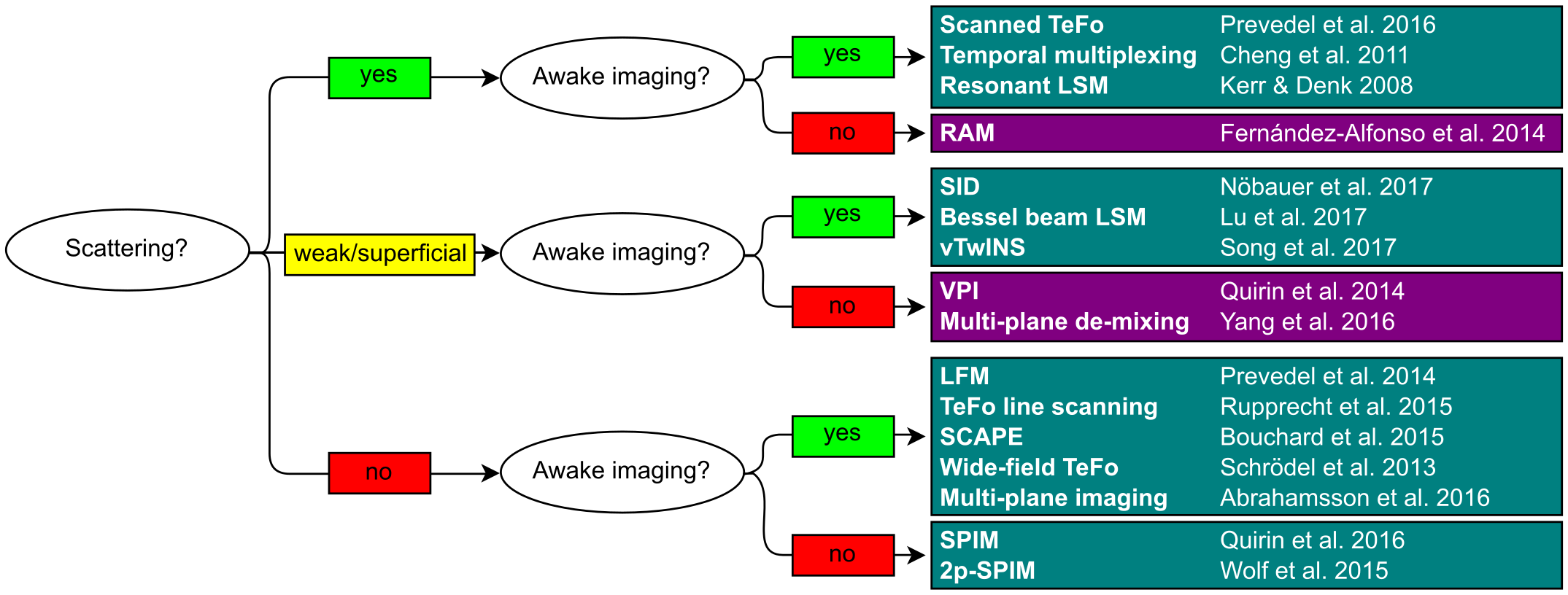}
\begin{figure*}[!htbp]
\centering \makeatletter\IfFileExists{0bc8781b-4931-4f3e-87b3-6890f6b4a3d4-ufig5_guide.png}{\includegraphics{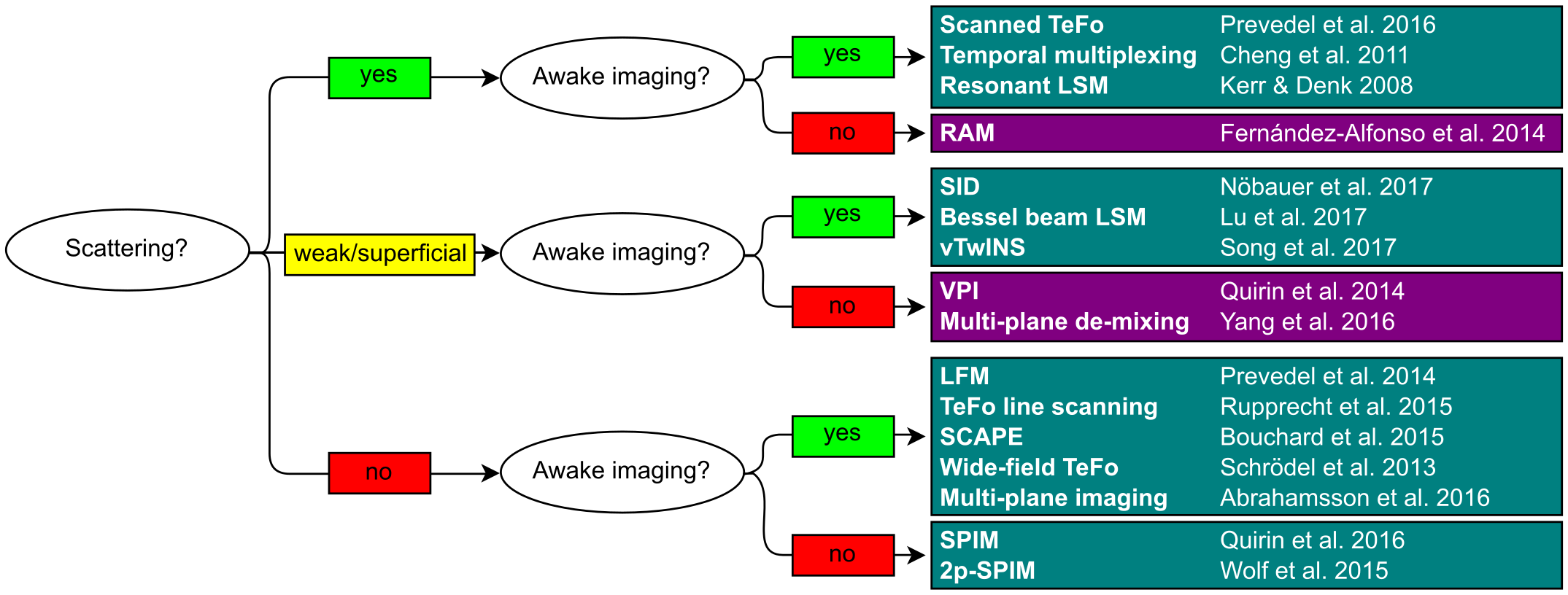}}{}
\makeatother 
\caption{{Selection guide for Ca\ensuremath{^{2+}} imaging techniques for different sample conditions and biological applications. Methods are categorized regarding their requirement for scattering compatibility and suitability for awake imaging which includes the susceptibility to motion artifacts. Example references for the techniques are provided.}}
\label{figure-fa93f91c04e3f8c0da2941577f355cbc}
\end{figure*}
\egroup
Figure 5 provides a summary of the most widely used microscopy techniques and their corresponding areas of application. For instance, unbiased imaging methods should be chosen for in vivo applications where significant sample motion is to be expected such as in freely behaving or awake head-fixed animals performing a motor tasks. In contrast, when sample motion is limited or knowledge about neuronal positions can be readily obtained, prior-based sampling is preferred as it reduces the number of sampled voxels and thereby increases the acquisition speed or the imaging volume. Another factor, as briefly discussed above, is the degree of tissue scattering and recording depth. In general, sequential methods, in particular 2p or multi-photon excitation, are preferred in scattering tissue at the cost of reduced acquisition speed. Methods using array detectors such as cameras can be faster but are limited by scattering and therefore more suited for non- or weakly scattering samples. LFM-based methods in combination with computational source extraction can considerably push the reachable depths for camera-based modalities in scattering tissue (N{\"{o}}bauer et al. 2017).  
    
\section{\textbf{Future developments}}
From the above discussions, it becomes clear that there is a unique set of conditions and parameter space, within which each of the microscopy techniques operate optimally. Thus, generally, hybrid methods can be expected to cover a broader application space, since they have the potential to combine advantages from multiple complementary techniques. Moreover, there are several frontiers for hardware improvement which mutually contribute to the ability to image larger volumes at faster acquisition rates. In the following, we will briefly discuss recent progress for faster scanners in LSM, cameras for parallel recording modalities and laser technology. 

A major limitation of the acquisition rates in LSM methods is speed of the beam scanning. When using mechanical scanning, the inertia is usually the limiting factor. This becomes particularly limiting in the case of axial scanning by moving the microscope objective. Thus, a strategy when using mechanical scanning is to reduce inertia by minimizing moving masses. For instance, axial scanning of a heavy microscope objective can be prevented by employing remote scanning approaches (Botcherby et al. 2008), based on scanning a light-weighted mirror via a voice coil (Rupprecht et al. 2016, Sofroniew et al. 2016), by using electro-tunable lenses (Chen et al. 2016, Grewe et al. 2011), electro-wetting lenses, or ultrasound (TAG) lenses (Kong et al. 2015). One way to increase the lateral scanning speed is to replace the commonly used combination of a resonant galvanometer scanner, which can achieve \textasciitilde 16 kHz line scan rates, and a galvanometer scanner with polygonal scanners (Bouchard et al. 2015, Li et al. 2017), or a combination of two resonant scanners that have a fixed phase relation and can be used to scan on a Lissajous trajectory (Newman et al. 2015). A variety of other technologies have been used for either lateral or axial scanning, including AODs (Grewe et al. 2010), electro-optic deflectors (Schneider et al. 2015), and all-optical scanning (Wu et al. 2017). In the case of parallel recording, ever-faster sensor arrays with larger number of pixels, faster read-out speed, and better light sensitivity are becoming available. Progress in these technical areas will further catalyze upscaling and the development of novel optical techniques.

The possibility to record at increasing volume rates over larger FOVs at sufficient SNR goes hand in hand with progress in the illumination sources. We believe that novel laser sources that can provide high pulse energy with adjustable repetition rates will be transformative as they allow to flexibly adapt the light source to the design of the imaging system for various applications and sample conditions (Prevedel et al. 2016). Fiber lasers or solid-state lasers at fixed wavelengths optimized for the wavelengths of well-engineered fluorophores will prove to be more useful than tunable titanium sapphire lasers. Lasers based on phenomena such as soliton self-frequency shift (Gruner-Nielsen et al. 2010, Zhu et al. 2013) are starting to become commercially available and will give access to a broader range of more wavelengths {\textendash} especially in the infrared. Ultimately, all high-speed Ca\ensuremath{^{2+}} imaging techniques will challenge fundamental limits such as energy dissipation (Marblestone et al. 2013) and constrains given by the properties of the currently used fluorescent probes. Progress in imaging techniques will also incentivize further developments in the probe development. We expect that together they will further push the boundaries of current possibilities for manipulating and recoding neuronal network activity from large population at high speed and resolution.

\section*{Acknowledgements}
We would like to acknowledge support from the National Institute of Neurological Disorders and Stroke of the US National Institutes of Health under Award Number U01NS103488 and U01NS094263-01 as well as The Kavli Foundation. S.W. has been supported by a Leon Levy Fellowship at The Rockefeller University.

Posted with permission from the Annual Review of Neuroscience, Volume 41 \textcopyright  2018 by Annual Reviews, http://www.annualreviews.org/.

\section*{List of abbreviations}
2p: two-photon

3p: three-photon 

AOD: acousto-optical deflector

Ca\ensuremath{^{2+}}: calcium

fMRI: functional magnetic resonance imaging

FOV: field of view

GECI: Genetically encoded calcium indicator

GFP: green fluorescent protein

LFM: light-field microscopy

LSM: laser scanning microscopy

NA: numerical aperture

PSF: point spread function

RAM: random access microscopy

SNR: signal-to-noise ratio

SBR: signal-to-background ratio

SPIM: selective plane illumination microscopy

TeFo: temporal focusing

\section*{\textbf{References}}
Abrahamsson S, Ilic R, Wisniewski J, Mehl B, Yu L, et al. 2016. Multifocus microscopy with precise color multi-phase diffractive optics applied in functional neuronal imaging. \textit{Biomed. Opt. Express}. 7(3):855{\textendash}15

Agard D. 1984. Optical Sectioning Microscopy: Cellular Architecture in Three Dimensions. \textit{Annual Review of Biophysics and Biomolecular Structure}. 13(1):191{\textendash}219

Ahrens MB, Orger MB, Robson DN, Li JM, Keller PJ. 2013. Whole-brain functional imaging at cellular resolution using light-sheet microscopy. \textit{Nat Meth}. 10(5):413{\textendash}20

Aimon S, Katsuki T, Grosenick L, Broxton M, Deisseroth K, et al. 2017. Fast whole brain imaging in adult Drosophila during response to stimuli and behavior. \textit{bioRxiv}. 1{\textendash}34

Akerboom J, Chen TW, Wardill TJ, Tian L, Marvin JS, et al. 2012. Optimization of a GCaMP Calcium Indicator for Neural Activity Imaging. \textit{Journal of Neuroscience}. 32(40):13819{\textendash}40

Amir W, Carriles R, Hoover EE, Planchon TA, Durfee CG, Squier JA. 2007. Simultaneous imaging of multiple focal planes using a two-photon scanning microscope. \textit{Opt. Lett.} 32(12):1731{\textendash}34

Averbeck BB, Latham PE, Pouget A. 2006. Neural correlations, population coding and computation. \textit{Nat Rev Neurosci}. 7(5):358{\textendash}66

Baumgart E, Kubitscheck U. 2012. Scanned light sheet microscopy with confocal slit detection. \textit{Opt. Express}. 20(19):21805{\textendash}10

Berenyi A, Somogyvari Z, Nagy AJ, Roux L, Long JD, et al. 2014. Large-scale, high-density (up to 512 channels) recording of local circuits in behaving animals. \textit{Journal of Neurophysiology}. 111(5):1132{\textendash}49

Botcherby EJ, Ju\v{s}kaitis R, Booth MJ, Wilson T. 2008. An optical technique for remote focusing in microscopy. \textit{Optics Communications}. 281(4):880{\textendash}87

Botcherby EJ, Ju\v{s}kaitis R, Wilson T. 2006. Scanning two photon fluorescence microscopy with extended depth of field. \textit{Optics Communications}. 268(2):253{\textendash}60

Bouchard MB, Voleti V, Mendes CS, Lacefield C, Grueber WB, et al. 2015. Swept confocally-aligned planar excitation (SCAPE) microscopy for high-speed volumetric imaging of behaving organisms. \textit{Nature Photonics}. 9(2):113{\textendash}19

Boyden ES, Zhang F, Bamberg E, Nagel G, Deisseroth K. 2005. Millisecond-timescale, genetically targeted optical control of neural activity. \textit{Nat Neurosci}. 8(9):1263{\textendash}68

Braitenberg V, Sch{\"{u}}z A. 1998. \textit{Cortex: Statistics and Geometry of Neuronal Connectivity}. Berlin, Heidelberg: Springer Berlin Heidelberg

Broxton M, Grosenick L, Yang S, Cohen N, Andalman A, et al. 2013. Wave optics theory and 3-D deconvolution for the light field microscope. \textit{Opt. Express}. 21(21):25418{\textendash}22

Buzsaki G. 2004. Large-scale recording of neuronal ensembles. \textit{Nat Neurosci}. 7(5):446{\textendash}51

Buzsaki G, Mizuseki K. 2014. The log-dynamic brain: how skewed distributions affect network operations. \textit{Nature Publishing Group}. 15(4):264{\textendash}78

Chamberland S, Yang HH, Pan MM, Evans SW, Guan S, et al. 2017. Fast two-photon imaging of subcellular voltage dynamics in neuronal tissue with genetically encoded indicators. \textit{eLIFE}. 1{\textendash}76

Chan KY, Jang MJ, Yoo BB, Greenbaum A, Ravi N, et al. 2017. Engineered AAVs for efficient noninvasive gene delivery to the central and peripheral nervous systems. \textit{Nat Neurosci}. 20(8):1172{\textendash}79

Chen JL, Voigt FF, Javadzadeh M, Krueppel R, Helmchen F. 2016. Long-range population dynamics of anatomically defined neocortical networks. \textit{eLIFE}. 1{\textendash}26

Chen T-W, Wardill TJ, Sun Y, Pulver SR, Renninger SL, et al. 2013. Ultrasensitive fluorescent proteins for imaging neuronal activity. \textit{Nature}. 499(7458):295{\textendash}300

Cheng A, Gon\c{c}alves JT, Golshani P, Arisaka K, Portera-Cailliau C. 2011. Simultaneous two-photon calcium imaging at different depths with spatiotemporal multiplexing. \textit{Nat Meth}. 8(2):139{\textendash}42

Cong L, Wang Z, Chai Y, Hang W, Shang C, et al. 2017. Rapid whole brain imaging of neural activity in freely behaving larval zebrafish (Danio rerio). 1{\textendash}55

Conkey DB, Caravaca-Aguirre AM, Piestun R. 2012. High-speed scattering medium characterization with application to focusing light through turbid media. \textit{Opt. Express}. 20(2):1733{\textendash}38

Denk W, Strickler JH, Webb WW. 1990. Two-Photon Laser Scanning Fluorescence Microscopy. \textit{Science}. 1{\textendash}5

Dittgen T, Nimmerjahn A, Komai S, Licznerski P, Waters J, et al. 2004. Lentivirus-based genetic manipulations of cortical neurons and their optical and electrophysiological monitoring in vivo. \textit{Proceedings of the National Academy of Sciences}. 101(52):18206{\textendash}11

Fern\'{a}ndez-Alfonso T, Nadella KMNS, Iacaruso MF, Pichler B, Ro\v{s} H, et al. 2014. Monitoring synaptic and neuronal activity in 3D with synthetic and genetic indicators using a compact acousto-optic lens two-photon microscope. \textit{Journal of Neuroscience Methods}. 222:69{\textendash}81

Foust KD, Nurre E, Montgomery CL, Hernandez A, Chan CM, Kaspar BK. 2008. Intravascular AAV9 preferentially targets neonatal neurons and adult astrocytes. \textit{Nature Biotechnology}. 27(1):59{\textendash}65

Friedrich J, Yang W, Soudry D, Mu Y, Ahrens MB, et al. 2017. Multi-scale approaches for high-speed imaging and analysis of large neural populations. \textit{PLoS Comput Biol}. 13(8):e1005685{\textendash}24

Gouwens NW, Wilson RI. 2009. Signal Propagation in Drosophila Central Neurons. \textit{Journal of Neuroscience}. 29(19):6239{\textendash}49

Grewe BF, Langer D, Kasper H, Kampa BM, Helmchen F. 2010. High-speed in vivo calcium imaging reveals neuronal network activity with near-millisecond precision. \textit{Nat Meth}. 7(5):399{\textendash}405

Grewe BF, Voigt FF, van Hoff M, Helmchen F. 2011. Fast two-layer two-photon imaging of neuronal cell populations using an electrically tunable lens. \textit{Biomed. Opt. Express}. 1{\textendash}12

Grienberger C, Konnerth A. 2012. Imaging Calcium in Neurons. \textit{Neuron}. 73(5):862{\textendash}85

Grinvald A, Salzberg BM, Lev-Ram V, Hildesheim R. 1987. Optical recording of synaptic potentials from processes of single neurons using intracellular potentiometric dyes. \textit{Biophysj}. 51(4):643{\textendash}51

Gruner-Nielsen L, Jakobsen D, Jespersen KG, P\'{a}lsd\'{o}ttir B. 2010. A stretcher fiber for use in fs chirped pulse Yb amplifiers. \textit{Opt. Express}. 18(4):3768{\textendash}6

Grynkiewicz G, Poenie M, Tsien RY. 1985. New Generation of Ca2+ Indicators with Greatly Improved Fluorescence Properties. \textit{The Journal of Biological Chemistry}. 3440

Harris KD, Quiroga RQ, Freeman J, Smith SL. 2016. Improving data quality in neuronal population recordings. \textit{Nat Neurosci}. 19(9):1165{\textendash}74

Herculano-Houzel S, Collins CE, Wong P, Kaas JH. 2007. Cellular scaling rules for primate brains. \textit{Proceedings of the National Academy of Sciences}. 104(9):3562{\textendash}67

Herculano-Houzel S, Mota B, Lent R. 2006. Cellular scaling rules for rodent brains. \textit{Proceedings of the National Academy of Sciences}. 103(32):12138{\textendash}43

Hilgen G, Sorbaro M, Pirmoradian S, Muthmann J-O, Kepiro IE, et al. 2017. Unsupervised Spike Sorting for Large-Scale, High- Density Multielectrode Arrays. \textit{CellReports}. 18(10):2521{\textendash}32

Hochbaum DR, Zhao Y, Farhi SL, Klapoetke N, Werley CA, et al. 2014. All-optical electrophysiology in mammalian neurons using engineered microbial rhodopsins. \textit{Nat Meth}. 11(8):825{\textendash}33

Hopt A, Neher E. 2001. Highly Nonlinear Photodamage in Two-Photon Fluorescence Microscopy. \textit{Biophysical Journal}. 1{\textendash}8

Horstmeyer R, Ruan H, Yang C. 2015. Guidestar-assisted wavefront-shaping methods for focusing light into biological tissue. \textit{Nature Photonics}. 1{\textendash}9

Horton NG, Wang K, Kobat D, Clark CG, Wise FW, et al. 2013. SI: In vivo three-photon microscopy of subcortical structures within an intact mouse brain. \textit{Nature Photonics}. 7(3):205{\textendash}9

Horton NG, Xu C. 2015. Dispersion compensation in three-photon fluorescence microscopy at 1,700 nm. \textit{Biomed. Opt. Express}. 6(4):1392{\textendash}96

Howarth C, Gleeson P, Attwell D. 2012. Updated energy budgets for neural computation in the neocortex and cerebellum. 32(7):1222{\textendash}32

Huettel SA, Song AW, McCarthy GJ. 2008. \textit{Functional Magnetic Resonance Imaging}. Sinauer Associates. Second Edition ed.

Huisken J, Swoger J, Del Bene F, Wittbrodt J, Stelzer EHK. 2004. Optical Sectioning Deep Inside Live Embryos by Selective Plane Illumination Microscopy. \textit{Science}. 305:1{\textendash}18

Insel TR, Landis SC, Collins FS. 2013. The NIH BRAIN Initiative. \textit{Science}. 340:687

Ji G, Feldman ME, Deng K-Y, Greene KS, Wilson J, et al. 2004. Ca 2+-sensing Transgenic Mice. \textit{JOURNAL OF BIOLOGICAL CHEMISTRY}. 279(20):21461{\textendash}68

Kalmbach AS, Waters J. 2012. Brain surface temperature under a craniotomy. \textit{Journal of Neurophysiology}. 108(11):3138{\textendash}46

Kato S, Kaplan HS, Schr{\"{o}}del T, Skora S, Lindsay TH, et al. 2015. Global Brain Dynamics Embed the Motor Command Sequence of Caenorhabditis elegans. \textit{Cell}. 163(3):656{\textendash}69

Katona G, Szalay G, Ma\'{a}k P, Kaszas A, Veress M, et al. 2012. Fast two-photon in vivo imaging with three-dimensional random-access scanning in large tissue volumes. \textit{Nat Meth}. 9(2):201{\textendash}8

Katz O, Small E, Guan Y, Silberberg Y. 2014. Noninvasive nonlinear focusing and imaging through strongly scattering turbid layers. \textit{Optica}. 1(3):170{\textendash}75

Kerr JND, Denk W. 2008. Imaging in vivo: watching the brain in action. \textit{Nat Rev Neurosci}. 9(3):195{\textendash}205

Koester HJ, Baur D, Uhl R, Hell SW. 1999. Fluorescence Imaging with Pico- and Femtosecond Two-Photon Excitation: Signal and Photodamage. \textit{Biophysical Journal}. 1{\textendash}11

Kong L, Tang J, Little JP, Yu Y, L{\"{a}}mmermann T, et al. 2015. Continuous volumetric imaging via an optical phase-locked ultrasound lens. \textit{Nat Meth}. 12(8):759{\textendash}62

Kovacevic N. 2004. A Three-dimensional MRI Atlas of the Mouse Brain with Estimates of the Average and Variability. \textit{Cerebral Cortex}. 15(5):639{\textendash}45

Kr{\"{u}}ger J. 2005. Simultaneous individual recordings from many cerebral neurons: Techniques and results. In \textit{Reviews of Physiology, Biochemistry and Pharmacology, Volume 98}, Vol. 98, pp. 177{\textendash}233. Berlin, Heidelberg: Springer Berlin Heidelberg

Kumar S, Wilding D, Sikkel MB, Lyon AR, MacLeod KT, Dunsby C. 2011. High-speed 2D and 3D fluorescence microscopy of cardiac myocytes. \textit{Opt. Express}. 19(15):13839{\textendash}39

Lai P, Wang L, Tay JW, Wang LV. 2015. Photoacoustically guided wavefront shaping for enhanced optical focusing in scattering media. \textit{Nature Photonics}. 9(2):126{\textendash}32

Lemon WC, ckendorf BHO, McDole K, Branson K, Freeman J, et al. 2015. Whole-central nervous system functional imaging in larval Drosophila. \textit{Nature Communications}. 6:1{\textendash}16

Levoy M, Ng R, Adams A, Footer M, Horowitz M. 2006. Light Field Microscopy. \textit{AMC}. 1{\textendash}11

Li YX, Gautam V, Br{\"{u}}stle A, Cockburn IA, Daria VR, et al. 2017. Flexible polygon-mirror based laser scanning microscope platform for multiphoton in-vivoimaging. \textit{J. Biophoton.} 2:143{\textendash}12

Lin MZ, Schnitzer MJ. 2016. Genetically encoded indicators of neuronal activity. \textit{Nat Neurosci}. 19(9):1142{\textendash}53

Lippmann MG. 1908. \'{E}preuves r\'{e}versibles. Photographies int\'{e}grales. \textit{Computes Rendos delAcademie des Sciences}. 146:446{\textendash}51

Liu H-Y, Jonas E, Tian L, Zhong J, Recht B, Waller L. 2015. 3D imaging in volumetric scattering media using phase-space measurements. \textit{Opt. Express}. 23(11):14461{\textendash}11

Lu R, Sun W, Liang Y, Kerlin A, Bierfeld J, et al. 2017. Video-rate volumetric functional imaging of the brain at synaptic resolution. \textit{Nat Neurosci}

Mao T, O'Connor DH, Scheuss V, Nakai J, Svoboda K. 2008. Characterization and Subcellular Targeting of GCaMP-Type Genetically-Encoded Calcium Indicators. \textit{PLoS ONE}. 3(3):e1796{\textendash}10

Marblestone AH, Zamft BM, Maguire YG, Shapiro MG, Cybulski TR, et al. 2013. Physical principles for scalable neural recording. \textit{Frontiers in Computational Neuroscience}. 7:

Martin, Jonathan D Thiessen, Laryssa M Kurjewicz, Shelley L Germscheid, Allan J Turner, et al. 2010. Longitudinal Brain Size Measurements in APP/PS1 Transgenic Mice. \textit{MRI}. 19{\textendash}8

Miyawaki A, Llopis J, Heim R, McCaffery JM, Adams JA, et al. 1997. Fluorescent indicators for Ca2+based on green fluorescent proteins and calmodulin. \textit{Nature}. 388(6645):882{\textendash}87

Mukamel EA, Nimmerjahn A, Schnitzer MJ. 2009. Automated Analysis of Cellular Signals from Large-Scale Calcium Imaging Data. \textit{Neuron}. 63(6):747{\textendash}60

Nagai T, Yamada S, Tominaga T, Ichikawa M, Miyawaki A. 2004. Expanded dynamic range of fluorescent indicators for Ca2\_ by circularly permuted yellow fluorescent proteins. \textit{Proceedings of the National Academy of Sciences}. 1{\textendash}6

Nagel G, Ollig D, Fuhrmann M, Kateriya S, Musti AM, et al. 2002. Channelrhodopsin-1: A Light-Gated Proton Channel in Green Algae. \textit{Science}. 296(5577):2395{\textendash}98

Nakai J, Ohkura M, Imoto K. 2001. A high signal-to-noise Ca2+ probe composed of a single green fluorescent protein. \textit{Nature Biotechnology}. 19:137

Naumann EA, Kampff AR, Prober DA, Schier AF, Engert F. 2010. Monitoring neural activity with bioluminescence during natural behavior. \textit{Nature Publishing Group}. 13(4):513{\textendash}20

Newman JA, Sullivan SZ, Muir RD, Sreehari S, Bouman CA, Simpson GJ. 2015. Multi-channel beam-scanning imaging at kHz frame rates by Lissajous trajectory microscopy. \textit{SPIE BiOS}. 9330:933009{\textendash}8

N{\"{o}}bauer T, Skocek O, Pern\'{\i}a-Andrade AJ, Weilguny L, Traub FM, et al. 2017. Video rate volumetric Ca2+ imaging across cortex using seeded iterative demixing (SID) microscopy. \textit{Nature Publishing Group}. 1{\textendash}10

Ntziachristos V. 2010. Going deeper than microscopy: the optical imaging frontier in biology. \textit{Nature Publishing Group}. 7(8):603{\textendash}14

Ogawa S, Lee TM, Kay AR, Tank DW. 1990. Brain magnetic resonance imaging with contrast dependent on blood oxygenation. \textit{Proceedings of the National Academy of Sciences}. 87(24):9868{\textendash}72

Oron D, Tal E, Silberberg Y. 2005. Scanningless depth-resolved microscopy. \textit{Opt. Express}. 1{\textendash}9

Ouzounov DG, Wang T, Wang M, Feng DD, Horton NG, et al. 2017. In vivo three-photon imaging of activity of GCaMP6-labeled neurons deep in intact mouse brain. \textit{Nature Publishing Group}. 1{\textendash}5

Papadopoulos IN, Jouhanneau J-S, Poulet JFA, Judkewitz B. 2016. Scattering compensation by focus scanning holographic aberration probing (F-SHARP). \textit{Nature Photonics}. 1{\textendash}9

Pawley JB. 2006. \textit{Handbook of Biological Confocal Microscopy}. Springer. 3rd ed.

Pnevmatikakis EA, Soudry D, Gao Y, Machado TA, Merel J, et al. 2016. Simultaneous Denoising, Deconvolution, and Demixing of Calcium Imaging Data. \textit{Neuron}. 89(2):285{\textendash}99

Podgorski K, Ranganathan GN. 2016. Brain heating induced by near infrared lasers during multi-photon microscopy. \textit{Journal of Neurophysiology}. jn.00275.2016{\textendash}12

Prevedel R, Verhoef AJ, Weisenburger S, Pern\'{\i}a-Andrade AJ, Huang BS, et al. 2016. Fast volumetric calcium imaging across multiple cortical layers using sculpted light. \textit{Nat Meth}. 13(12):1021{\textendash}28

Prevedel R, Yoon Y-G, Hoffmann M, Pak N, Wetzstein G, et al. 2014. Simultaneous whole-animal 3D imaging of neuronal activity using light-field microscopy. \textit{Nat Meth}. 11(7):727{\textendash}30

Quirin S, Jackson J, Peterka DS, Yuste R. 2014. Simultaneous imaging of neural activity in three dimensions. \textit{Front. Neural. Circuits}. 8(65):413{\textendash}11

Quirin S, Vladimirov N, Yang C-T, Peterka DS, Yuste R, B Ahrens M. 2016. Calcium imaging of neural circuits with extended depth-of-field light-sheet microscopy. \textit{Opt. Lett.} 41(5):855{\textendash}4

Reddy GD, Kelleher K, Fink R, Saggau P. 2008. Three-dimensional random access multiphoton microscopy for functional imaging of neuronal activity. \textit{Nat Neurosci}. 11(6):713{\textendash}20

Rein K, Z{\"{o}}ckler M, Mader MT, Gr{\"{u}}bel C, Heisenberg M. 2002. The Drosophila Standard Brain. \textit{Current Biology}. 12(3):227{\textendash}31

Rey HG, Pedreira C, Quiroga RQ. 2015. Past, present and future of spike sorting techniques. \textit{Brain Research Bulletin}. 119(Part B):106{\textendash}17

Rupprecht P, Prendergast A, Wyart C, Friedrich RW. 2016. Remote z-scanning with a macroscopic voice coil motor for fast 3D multiphoton laser scanning microscopy. \textit{Biomed. Opt. Express}. 7(5):1656{\textendash}16

Rupprecht P, Prevedel R, Groessl F, Haubensak WE, Vaziri A. 2015. Optimizing and extending light-sculpting microscopy for fast functional imaging in neuroscience. \textit{Biomed. Opt. Express}. 6(2):353{\textendash}16

Sahin B, Aslan H, Unal B, Canan S, Bilgic S, et al. 2001. Brain volumes of the lamb, rat and bird do not show hemispheric asymmetry: a stereological study. \textit{Image Analysis Stereology}. 20(1):1{\textendash}5

Salzberg BM, Grinvald A, Cohen LB, Davila HV, Ross WN. 1977. Optical Recording of Neuronal Activity in an Invertebrate Central Nervous System: Simultaneous Monitoring of Several Neurons. \textit{Journal of Neurophysiology}

Santi PA. 2011. Light Sheet Fluorescence Microscopy. \textit{J Histochem Cytochem.} 59(2):129{\textendash}38

Schneider J, Zahn J, Maglione M, Sigrist SJ, Marquard J, et al. 2015. Ultrafast, temporally stochastic STED nanoscopy of millisecond dynamics. \textit{Nat Meth}. 12(9):827{\textendash}30

Schr{\"{o}}del T, Prevedel R, Aumayr K, Zimmer M, Vaziri A. 2013. Brain-wide 3D imaging of neuronal activity in Caenorhabditis elegans with sculpted light. \textit{Nat Meth}. 10(10):1013{\textendash}20

Shimogori T, Ogawa M. 2008. Gene application with in uteroelectroporation in mouse embryonic brain. \textit{Development, Growth \& Differentiation}. 50(6):499{\textendash}506

Shimomura O, Johnson FH, Saiga Y. 1962. Extraction, Purification and Properties of Aequorin, a Bioluminescent Protein from the Luminous Hydromedusan,Aequorea. \textit{J. Cell. Comp. Physiol.} 59(3):223{\textendash}39

Simpson JH. 2009. Mapping and Manipulating Neural Circuits in the Fly Brain. In \textit{Genetic Dissection of Neural Circuits and Behavior}, Vol. 65, pp. 79{\textendash}143. Elsevier Inc. 1st ed.

So P, Dong CY, Masters B, Berland K. 2000. Two-photon excitation fluorescence microscopy. \textit{Annu. Rev. Neurosci.}

Sofroniew NJ, Flickinger D, King J, Svoboda K. 2016. A large field of view two-photon mesoscope with subcellular resolution for in vivo imaging. \textit{eLIFE}. 1{\textendash}20

Song A, Charles AS, Koay SA, Gauthier JL, Thiberge SY, et al. 2017. Volumetric two-photon imaging of neurons using stereoscopy (vTwINS). \textit{Nature Publishing Group}. 14(4):420{\textendash}26

Spiecker H. 2011. Verfahren und Anordnung zur Mikroskopie. \textit{DE102010013223A1}

Stirman JN, Smith IT, Kudenov MW, Smith SL. 2016. Wide field-of-view, multi-region, two-photon imaging of neuronal activity in the mammalian brain. \textit{Nature Biotechnology}. 34(8):865{\textendash}70

Tallini YN, Ohkura M, Choi BR, Ji G, Imoto K, et al. 2006. Imaging cellular signals in the heart in vivo: Cardiac expression of the high-signal Ca2+ indicator GCaMP2. \textit{Proceedings of the National Academy of Sciences}. 103(12):4753{\textendash}58

Th\'{e}riault G, Cottet M, Castonguay A, McCarthy N, De Koninck Y. 2014. Extended two-photon microscopy in live samples with Bessel beams: steadier focus, faster volume scans, and simpler stereoscopic imaging. \textit{Frontiers in Cellular Neuroscience}. 1{\textendash}11

Tian L, Hires SA, Mao T, Huber D, Chiappe ME, et al. 2009. Imaging neural activity in worms, flies and mice with improved GCaMP calcium indicators. \textit{Nature Publishing Group}. 6(12):875{\textendash}81

Trebino R, Zeek E. 2000. Ultrashort Laser Pulses. In \textit{Frequency-Resolved Optical Gating: the Measurement of Ultrashort Laser Pulses}, pp. 11{\textendash}35. Boston, MA: Springer US

Truong TV, Supatto W, Koos DS, Choi JM, Fraser SE. 2011. Deep and fast live imaging with two-photon scanned light-sheet microscopy. \textit{Nat Meth}. 8(9):757{\textendash}60

Vaziri A, Emiliani V. 2012. Reshaping the optical dimension in optogenetics. \textit{Current Opinion in Neurobiology}. 22(1):128{\textendash}37

Vellekoop IM, Mosk AP. 2007. Focusing coherent light through opaque strongly scattering media. \textit{Opt. Lett.} 32(16):2309{\textendash}11

Voie AH, Burns DH, Spelman FA. 1993. Orthogonal-plane fluorescence optical sectioning: Three-dimensional imaging of macroscopic biological specimens. \textit{Journal of Microscopy}. 170(3):229{\textendash}36

Wang T, Ouzounov D, Wang M, Xu C. 2017. Quantitative Comparison of Two-photon and Three-photon Activity Imaging of GCaMP6s-labeled Neurons in vivo in the Mouse Brain. \textit{Brain}, p. BrM4B.4. Washington, D.C.: OSA

Weisenburger S, Sandoghdar V. 2015. Light microscopy: an ongoing contemporary revolution. \textit{Contemporary Physics}. 56(2):123{\textendash}43

White JG, Southgate E, Thomson JN, Brenner S. 1986. The Structure of the Nervous System of the Nematode Caenorhabditis elegans. \textit{Phil. Trans. R. Soc. B}. 314(1165):1{\textendash}340

Wolf S, Supatto W, Debr\'{e}geas G, Mahou P, Kruglik SG, et al. 2015. Whole-brain functional imaging with two-photon light-sheet microscopy. \textit{Nature Publishing Group}. 12(5):379{\textendash}80

Wu J, Tang AHL, Mok ATY, Yan W, Chan GCF, et al. 2017. Multi-MHz laser-scanning single-cell fluorescence microscopy by spatiotemporally encoded virtual source array. \textit{Biomed. Opt. Express}. 8(9):4160{\textendash}12

Xu Y, Zou P, Cohen AE. 2017. Voltage imaging with genetically encoded indicators. \textit{Current Opinion in Chemical Biology}. 39:1{\textendash}10

Yang W, Miller J-EK, Carrillo-Reid L, Pnevmatikakis E, Paninski L, et al. 2016. Simultaneous Multi-plane Imaging of Neural Circuits. \textit{Neuron}. 89(2):269{\textendash}84

Yuste R, Bargmann C. 2017. Toward a Global BRAIN Initiative. \textit{Cell}. 168(6):956{\textendash}59

Zariwala HA, Borghuis BG, Hoogland TM, Madisen L, Tian L, et al. 2012. A Cre-Dependent GCaMP3 Reporter Mouse for Neuronal Imaging In Vivo. \textit{Journal of Neuroscience}. 32(9):3131{\textendash}41

Zhang F, Aravanis AM, Adamantidis A, de Lecea L, Deisseroth K. 2007. Circuit-breakers: optical technologies for probing neural signals and systems. \textit{Nat Rev Neurosci}. 8(9):732{\textendash}32

Zhao M, Zhang H, Li Y, Ashok A, Liang R, et al. 2014. Cellular imaging of deep organ using two-photon Bessel light-sheet nonlinear structured illumination microscopy. \textit{Biomed. Opt. Express}. 5(5):1296{\textendash}13

Zhu G, van Howe J, Durst M, Zipfel W, Xu C. 2005. Simultaneous spatial and temporal focusing of femtosecond pulses. \textit{Opt. Express}. 1{\textendash}7

Zhu L, Verhoef AJ, Jespersen KG, Kalashnikov VL, Gr{\"{u}}ner-Nielsen L, et al. 2013. Generation of high fidelity 62-fs, 7-nJ pulses at 1035 nm from a net normal-dispersion Yb-fiber laser with anomalous dispersion higher-order-mode fiber. \textit{Opt. Express}. 21(14):16255{\textendash}58

\bibliographystyle{elsarticle-num}

\bibliography{\jobname}

\end{document}